\documentclass[proof]{pasj01}

\usepackage{natbib}
\usepackage{bm}
\usepackage{subfigure}

\newcommand{\MFOF}{M_{\rm FOF}}

\newcommand{\Mvir}{M_{\rm vir}}
\newcommand{\Mel}{M_{\rm ellipsoid}}

\begin{document}

\title{Evolution and Statistics of Non-Sphericity of Dark Matter Halos from Cosmological N-Body Simulation}

\author{
Daichi Suto \altaffilmark{1}, 
Tetsu Kitayama \altaffilmark{2}, 
Takahiro Nishimichi \altaffilmark{3,4},
Shin Sasaki \altaffilmark{5},
Yasushi Suto \altaffilmark{1,6}
}

\email{daichi@utap.phys.s.u-tokyo.ac.jp}

\altaffiltext{1}{Department of Physics, The University of Tokyo, Tokyo 113-0033, Japan}
\altaffiltext{2}{Department of Physics, Toho University,  Funabashi, Chiba 274-8510, Japan}
\altaffiltext{3}{Kavli Institute for the Physics and Mathematics of the Universe (WPI), The University of Tokyo Institutes for Advanced Study, The University of Tokyo, 5-1-5 Kashiwanoha, Kashiwa 277-8583, Japan}
\altaffiltext{4}{CREST, JST, 4-1-8 Honcho, Kawaguchi, Saitama, 332-0012, Japan}
\altaffiltext{5}{Department of Physics, Tokyo Metropolitan University, Hachioji, Tokyo 192-0397, Japan}
\altaffiltext{6}{Research Center for the Early Universe, School of Science, The University of Tokyo, Tokyo 113-0033, Japan}
\KeyWords{Cosmology: dark matter; large-scale structure of Universe; Galaxies: clusters: general}

\maketitle

\begin{abstract}

We revisit the non-sphericity of cluster-mass scale halos from cosmological N-body simulation on the basis of triaxial modelling. In order to understand the difference between the simulation results and the conventional ellipsoidal collapse model (EC), we first consider the evolution of {\it individual} simulated halos. The major difference between EC and the simulation becomes appreciable after the turn-around epoch. Moreover, it is sensitive to the individual evolution history of each halo. Despite such strong dependence on individual halos, the resulting non-sphericity of halos exhibits weak but robust mass dependence in a {\it statistical} fashion; massive halos are more spherical up to the turn-around, but gradually become less spherical by $z=0$. This is clearly inconsistent with the EC prediction; massive halos are usually more spherical. In addition, at $z=0$, inner regions of the simulated halos are less spherical than outer regions, i.e., the density distribution inside the halos is highly inhomogeneous and therefore not self-similar (concentric ellipsoids with the same axis ratio and orientation). This is also inconsistent with the homogeneous density distribution that is commonly assumed in EC. Since most of previous fitting formulae for the probability distribution function (PDF) of axis ratio of triaxial ellipsoids have been constructed under the self-similarity assumption, they are not accurate. Indeed, we compute the PDF of {\it projected} axis ratio $a_1/a_2$ directly from the simulation data {\it without} the self-similarity assumption, and find that it is very sensitive to the assumption. The latter needs to be carefully taken into account in direct comparison with observations, and therefore we provide an empirical fitting formula for the PDF of $a_1/a_2$. Our preliminary analysis suggests that the derived PDF of $a_1/a_2$ roughly agrees with the current weak-lensing observations. More importantly, the present results will be useful in future exploration of the non-sphericity of clusters in X-ray and optical observations.
\end{abstract}

\section{Introduction}

Dark matter halos serve as building blocks in the structure formation in the universe. While the spherical assumption for shapes of dark matter halos has been widely used both in theoretical and observational researches, a number of observations and cosmological simulations has exhibited clear signatures of the non-sphericity of dark matter halos.

It is conventionally accepted that the primordial density fluctuations in the early universe obey the Gaussian random field, and they have a definite statistical signature of the non-sphericity \citep{Doroshkevich70,Bardeen86}. One prescription for the non-spherical evolution of such primordial density fluctuations is given by the ellipsoidal collapse model \citep[hereafter EC;][]{White79,Bond96}. By taking account of the non-spherical evolution of halos, \cite{Sheth02} found that the mass function fits better the simulation results than that of spherical prediction based on the Press-Schechter theory \citep{Press74}.

The improved quality of observational data demands more accurate theoretical models of the non-sphericity of halos beyond the simple EC prediction. However, purely theoretical description of the non-spherical structure and evolution of dark matter halos is very difficult due to the non-linear evolution and complicated interactions among dark matter within the highly inhomogeneous density distribution. Therefore cosmological N-body simulations play key roles in investigating the non-sphericity of halos. In fact, many authors have recently studied the non-sphericity of halos extracted from cosmological N-body simulations \citep{Jing02,Ludlow11,Schneider12,Bryan13,Despali13,Despali14,Borzyszkowski14,Ludlow14,
Butsky15,Velliscig15,Bonamigo15,Vega16}.

In particular, \cite{Jing02} (hereafter JS02) modelled simulated halos by triaxial ellipsoids, and found that their minor-to-major axis ratio follows a universal probability distribution function (PDF) that depends on redshift and mass. \cite{Rossi11} found that the PDF of JS02 contradicts the prediction of EC with the Gaussian random initial conditions; while massive halos are more spherical in EC, they are less spherical in JS02. \cite{Bonamigo15} and \cite{Vega16} performed improved N-body simulations for halos with higher resolution in a wider mass range, and they confirmed that their results basically reproduce JS02, still in disagreement with EC.

Even from the existing observation data, the non-sphericity of halos is already detectable especially for galaxy clusters in X-ray and optical bands. Since observational analyses are based on the density distribution of halos {\it projected on the sky}, the projected (two-dimensional) non-sphericity is more relevant quantity than the three-dimensional one. In fact, \cite{Oguri03} (hereafter OLS03) calculated the PDF of projected axis ratio by integrating the PDF of JS02. Then they showed that the observed excess of gravitationally lensed arcs relative to the spherical model prediction can be reconciled by taking into account the effect of the non-sphericity of lensing halos. The PDF of the non-sphericity has been also observationally examined, although the observational uncertainty is large at this stage. The weak lensing study by \cite{Oguri10} showed that their 18 clusters have a PDF of ellipticity barely consistent with that proposed by JS02. The 70 X-ray clusters analyzed by \cite{Kawahara10} with the hydrostatic equilibrium assumption also produced a roughly consistent result with JS02. At this stage, the available data is limited and the observational uncertainty is large. In the near future, however, especially Subaru Hyper Suprime-Cam\footnote{www.naoj.org/Projects/HSC/} will provide us with a number of highly resolved lensing halos that are suitable for non-spherical analyses.

We emphasize that the PDF of JS02, and therefore that of OLS03 assume the self-similarity of the halo structure; the density distribution of halos is approximated by concentric ellipsoids with the same axis ratio and orientation. As JS02 have already indicated, however, the simulated halos are not necessarily self-similar. In order to fully utilize the data by future observations, an improved model of the projected axis ratio is quite important.

Therefore, our goal in this paper is to find an empirical fitting formula for the PDF of projected axis ratio of dark matter halos without assuming their self-similarity. For that purpose, we analyze 2004 halos extracted from N-body simulation, and approximate the density distribution around each simulated halo and its protohalos by triaxial ellipsoids. To better understand the non-spherical evolution of the halos, we first compare the evolution of individual halos with the EC prediction. Next we statistically examine the mass dependence and the radial profile of axis ratio of the simulated halos, and identify how and when the difference between EC and simulations emerges. Importantly, we find that the self-similarity assumption for halos adopted in the previous studies (JS02; OLS03) is not accurate, and that the prediction of the projected non-sphericity of dark matter halos is significantly affected by the assumption. Therefore we calculate the PDF of the projected axis ratio directly from the simulation data. This is exactly what we aim at in this paper.

The rest of this paper is organized as follows. Section 2 describes our N-body simulation and how to follow the evolution of our simulated halos. We compare the evolution of individual halos with EC in Section 3. Section 4 discusses the statistical evolution of halo non-sphericity. In Section 5, we present the PDF of axis ratio of halos in three-dimensional space, and its two-dimensional counterparts are constructed. Finally Section 6 summarizes this paper.

\section{Triaxial Modelling of Simulated Halos}
\subsection{N-body simulation}

Throughout this paper, we use cluster-scale halos identified from a cosmological N-body simulation. The details of the simulation and the halo-finding procedures are described in this subsection.

We start the simulation at $z=99$, where $N=1024^3$ particles distributed in a periodic cube with a side length of 360 $h^{-1}\mathrm{Mpc}$ (comoving). Their initial conditions are generated with a parallel code developed by \cite{Nishimichi09} and \cite{Valageas11}, which is based on the second-order Lagrangian perturbation theory \citep{Scoccimarro98,Crocce06}.

We employ the matter transfer function computed by a linear Boltzmann solver \texttt{CAMB} \citep{Lewis00} for a flat $\Lambda$CDM cosmology with the nine-year WMAP parameters \citep{Hinshaw13}; $\Omega_{\mathrm{m},0} = 0.279$, $h=0.7$, $n_\mathrm{s}=0.972$, and $\sigma_8=0.821$ are the current matter density in units of the critical density, the Hubble constant in units of $100\,\mathrm{km}\,\mathrm{s}^{-1}\mathrm{Mpc}^{-1}$, the scalar spectral index, and the amplitude of the density fluctuation (linearly extrapolated to the present) smoothed with a top-hat filter of radius 8 $h^{-1}\mathrm{Mpc}$, respectively. With the above parameters, mass of each simulation particle $m_{\rm particle}$ is $3.4\times10^9\,h^{-1}M_\odot$, which is sufficient to resolve massive halos ($\gtrsim 10^{14}h^{-1}M_\odot$) at $z=0$ .

The particle distribution is then evolved using a publicly available parallel cosmological $N$-body solver \texttt{Gadget2} \citep{Springel05}. The long-range gravitational force is computed on $2048^3$ mesh points based on the fast Fourier transform, while we rely on the tree algorithm with the softening length of $20\,h^{-1}\mathrm{kpc}$ on short range. We store snapshots at redshifts $z=49$, 9, 5, 4, 3, 2, 1.5, 1, 0.8, 0.6, 0.4, 0.2, 0.1 and 0. Halos at $z=0$ are identified using the friends-of-friends (\texttt{FOF}) algorithm \citep{Davis85} with the linking length of $0.159$ times the mean inter-particle separation in one dimension. This length is chosen so that the corresponding virial overdensity $\Delta_{\rm vir}$ matches 355.4 in units of the cosmic mean density at $z=0$, which is motivated by the spherical collapse model \citep{Gunn72,Gunn77,Peebles80}. Indeed, we confirmed that the total mass of the linked particles $\MFOF$ approximately corresponds to the virial mass $\Mvir$. We further apply the \texttt{SUBFIND} algorithm \citep{Springel01} implemented in \cite{Nishimichi14} for each FOF halo to identify substructures as well as unbound particles.

In this paper, we use the FOF halos with $\MFOF>6.25 \times 10^{13} h^{-1}M_\odot$, corresponding to the mass range of galaxy clusters which are well resolved in optical and X-ray observations. The total number of those halos is 2004.

\subsection{Morphology of FOF halos}

Before modelling the 2004 simulated halos by triaxial ellipsoids, we classify the halos by the amount of substructures. This is useful in understanding the extent to which the definition of the non-sphericity of the FOF halos is sensitive to the presence of substructures.

Due to the nature of the FOF algorithm, an FOF halo may comprise two or more prominent components. Such a halo tends to yield higher non-sphericity, which should be distinguished from a very elongated {\it single} structure.

According to the result of the {\tt SUBFIND} algorithm, we obtain the mass $M_i$ of the $i$-th most massive component for each FOF halo. The most massive component ($i=1$) is called the ``main halo'', and we call the other components ``substructures''. Since the shape of a cluster is expected to be sensitive to a few prominent substructures, we classify halos according to the values of $M_2/M_1$ and $M_3/M_1$, instead of the total mass fraction of substructures. We note that this classification is free from numerical resolutions because such big substructures are all well-resolved in the current simulation.

\begin{figure}[h]
\begin{center}
\subfigure{
\FigureFile(50mm,50mm){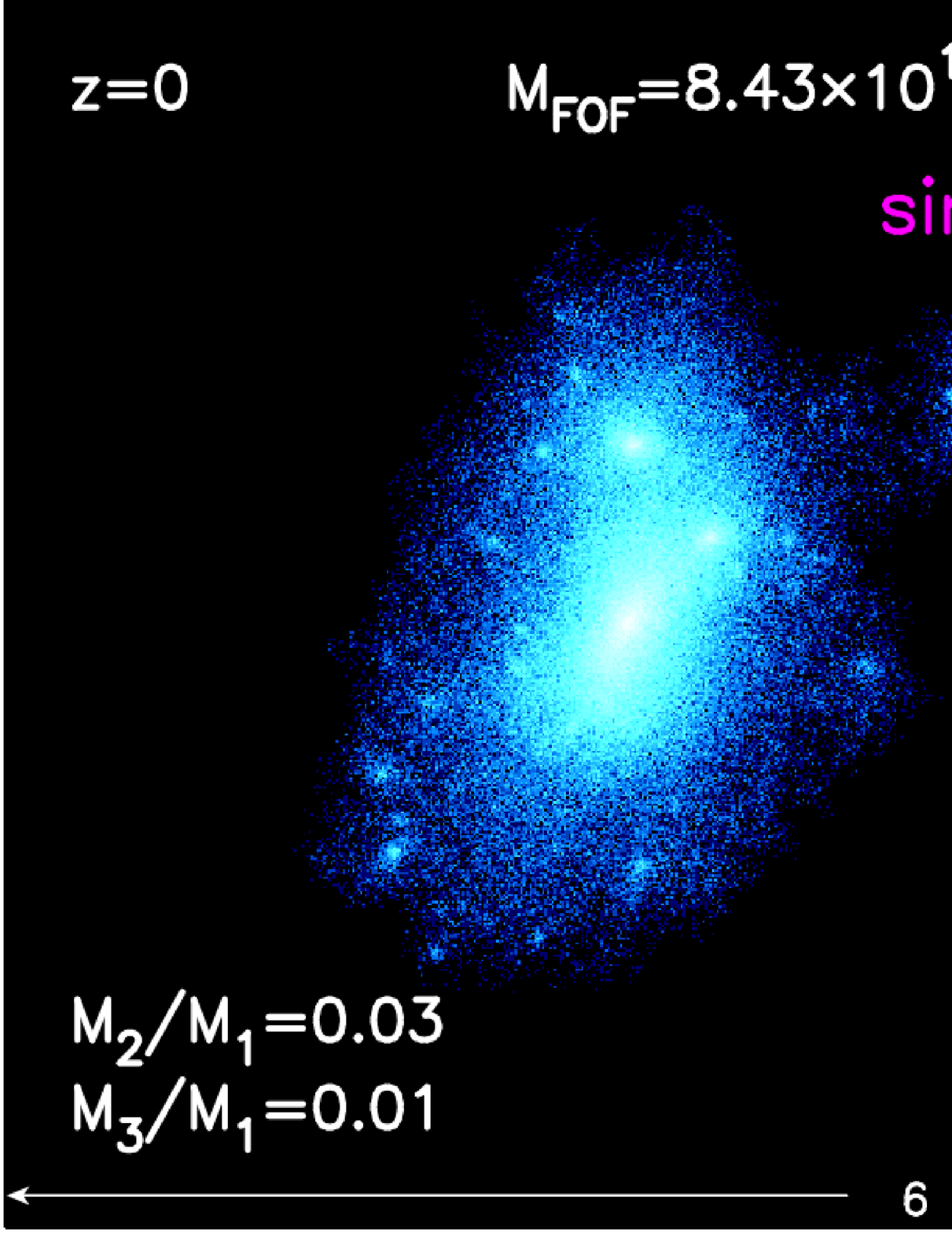}
\qquad
\FigureFile(50mm,50mm){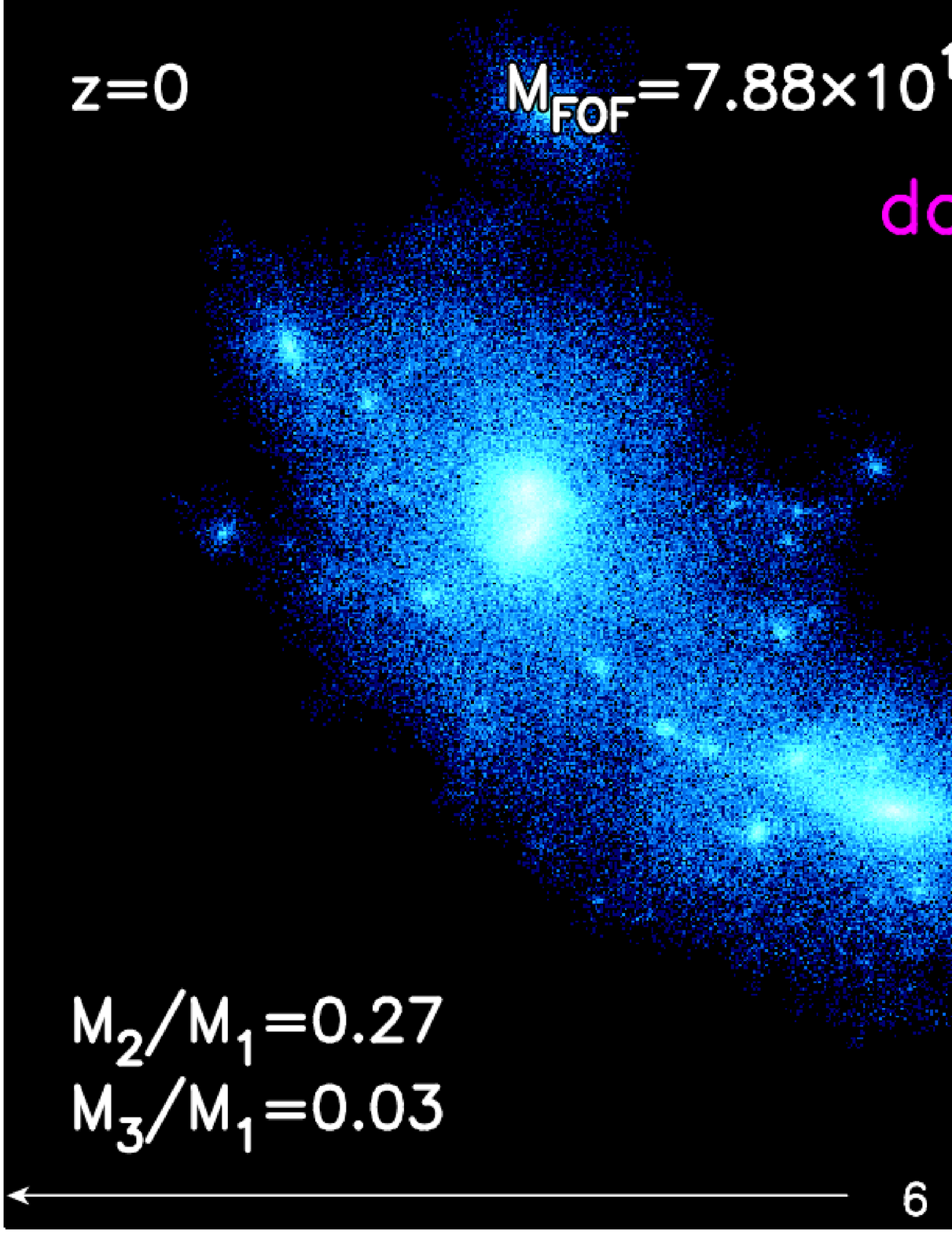}
}\\
\subfigure{
\FigureFile(50mm,50mm){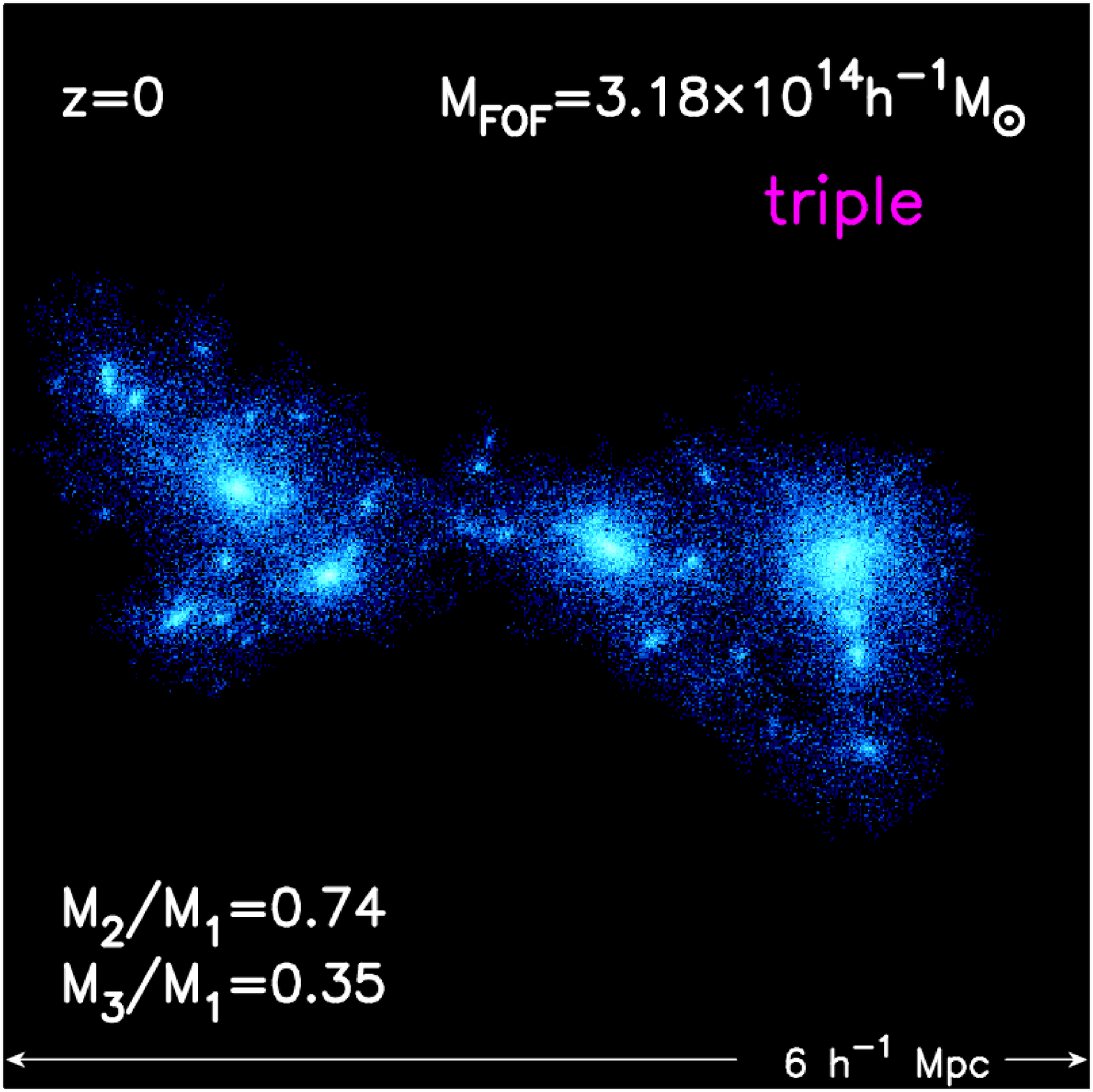}
\qquad
\FigureFile(50mm,50mm){frac.eps}
}
\end{center}
\caption{Examples of morphology of dark matter halos. The upper panels
and the lower-left panel show the FOF member particles of halos
comprising one, two and three major components in the cubic region 6
$h^{-1}$ Mpc a side around the halo. The FOF mass $\MFOF$ of each halo
is 8.43, 7.88 and 3.18 $\times10^{14}h^{-1}M_\odot$, respectively.  The
ratios of mass of the second and third massive components ($M_2$ and
$M_3$) compared to the mass of the most massive one (main halo) $M_1$
are also indicated in the three panels. The cumulative fractions of
$M_2/M_1$ and $M_3/M_1$ for our 2004 FOF halos are illustrated in the
lower-right panel. For example, $\sim$ 90 \% of the halos have
$M_2/M_1<0.2$. In this paper, we call the halos with $M_2/M_1>0.2$
``multiple-halos''. In this figure, the halos in the upper-right and
lower-left panels are multiple-halos.  } \label{class}
%
\begin{center}
\subfigure{
\FigureFile(50mm,50mm){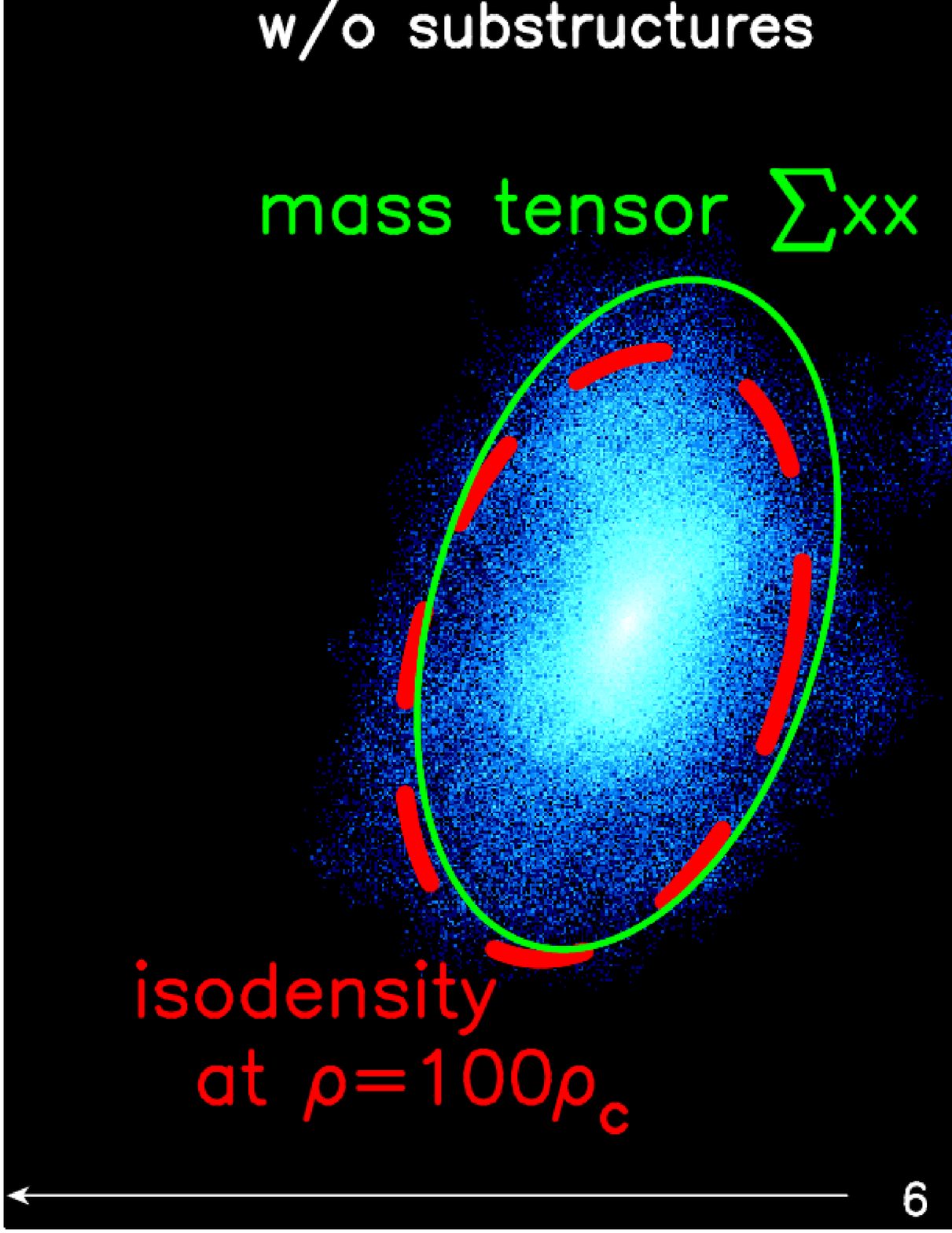}
}
\end{center}
\caption{Same as the top-left panel of Figure \ref{class}, but excluding
substructures. The projection of the ellipsoid fitted to the isodensity
surface at $\rho=100\rho_c$ is shown in the red dashed curve. The
ellipsoid enclosing the same mass as that inside the isodensity surface
($\Mel=6.25\times10^{14}h^{-1}M_\odot$) is determined by using the mass
tensor $I=\sum xx$ and its projection is plotted in the green solid
curve. The resulting two ellipsoids are similar; $A_1/A_3=0.57$
(isodensity surface) and $A_1/A_3=0.55$ (mass tensor).}  \label{mtcomp}
\end{figure}

The upper-left, upper-right and lower-left panels of Figure \ref{class} show the snapshots of FOF member particles of three halos with different morphology. The halo in the upper-left panel has very small values of $M_2/M_1$ and $M_3/M_1$, representing a single isolated structure. In contrast, the halo in the upper-right panel has relatively large $M_2/M_1$ and small $M_3/M_1$, corresponding to a ``double-structure''. The third halo in the lower-left panel has relatively large values both for $M_2/M_1$ and $M_3/M_1$, and is classified as a ``triple-structure''.

The lower-right panel of Figure \ref{class} indicates the cumulative fraction of the halos with a given threshold of $M_2/M_1$ or $M_3/M_1$. The majority of our halos have small $M_2/M_1$, and $M_3$ is substantially smaller than $M_2$. Thus the multiplicity of most of the halos can be characterized by the value of $M_2/M_1$. For later convenience, we set the threshold of $M_2/M_1=0.2$, and call a halo with $M_2/M_1<0.2$ a ``single-halo''. Also, a halo with $M_2/M_1>0.2$ is referred to as a ``multiple-halo''. Then the halos in the upper-right and the lower-left panels of Figure \ref{class} are multiple-halos. Such multiple-halos occupy approximately 10 \% of all the 2004 halos.

The threshold $M_2/M_1=0.2$ is somewhat arbitrary. According to the right panel of Figure \ref{class}, if we set the threshold by $M_2/M_1=0.1$, for example, $\sim$ 20 \% of our sample are classified as multiple-halos. As will be seen in the later sections, the choice of the threshold does not make a major difference in the main results of this paper.

Table 1 lists the number of the single- and multiple-halos out of our sample, corresponding to the threshold $M_2/M_1=0.2$.

\begin{table}
\begin{center}
\begin{tabular}{cc|ccc|c}
$M/10^{14}h^{-1}M_\odot$ & & $>$ 2.5 & 1.25 - 2.5 & 0.625 - 1.25 & total \\
\hline
single & $M_2/M_1<0.2$ & 171 &  429 & 1172 & 1772\\
\hline
multiple & $M_2/M_1>0.2$ & 32 & 68 & 132 & 232\\
\hline
total & & 203 & 497 & 1304 & 2004\\
\end{tabular}
\end{center}
\caption{The numbers of single- and multiple-halos, where $M_1$ and $M_2$ are the masses of the main halo and the most massive substructure.}
\end{table}

\subsection{Comparison of different methods of triaxial modelling: mass tensor vs. isodensity surface}

Throughout this paper, we approximate the density distribution of the simulated halos by a triaxial ellipsoid with the axis lengths $A_k$ ($k=1$, 2, 3). The boundary of the ellipsoid is described by
\begin{equation}
\left(\frac{x_1}{A_1}\right)^2+\left(\frac{x_2}{A_2}\right)^2+\left(\frac{x_3}{A_3}\right)^2=1,
\end{equation}
where $x_k$ ($k=1$, 2, 3) denotes the coordinate defined along the three axes with the origin set to the center of the ellipsoid.

One measure of the non-sphericity of each halo is the minor-to-major axis ratio $A_1/A_3$. The ellipticity $e$:
\begin{equation}
e=\frac{A_3-A_1}{2(A_1+A_2+A_3)}.
\label{eandp}
\end{equation}
is also used as an indicator of the non-sphericity in theliterature. Thus we consider both $A_1/A_3$ and $e$ in the following sections. For example, a sphere has $e=0$ and $A_1/A_3=1$. Also, for an ellipsoid with $A_1/A_3=0.5$, $0.1\le e\le0.125$, depending on $A_2$; $e$ is primarily determined by the deference between $A_1$ and $A_3$, whereas it also depends on $A_2$ (or “prolateness”).

The values of axis lengths $A_k$ are not constant for an entire simulated halo. Within the approximation of triaxial modelling, $A_k$ should be expressed as $A_k(\Mel)$, where $\Mel$ is the mass enclosed by the ellipsoid. Accordingly, $A_1/A_3$ and $e$ also depend on $\Mel$.

In this paper, we compute the axis lengths $A_k(\Mel)$ on the basis of the mass tensor $I_{\alpha\beta}$ (defined below) in an iterative fashion as follows. For a given set of $A_k(\Mel)$, we compute the mass tensor $I_{\alpha\beta}$:
\begin{equation}
I_{\alpha\beta}=\sum^N_{i=1} x^{(i)}_{\alpha}x^{(i)}_{\beta},
\label{Ixx}
\end{equation}
where $x^{(i)}_\alpha$ ($\alpha=1$, 2, 3) is the coordinate of the $i$-th particle along the three axes of the ellipsoid, and the summation is taken over the $N(=\Mel/m_{\rm particle})$ particles inside the ellipsoid. That mass tensor is now diagonalized and rotate the coordinate accordingly. The square root of the eigenvalues multiplied by some constant now become a new set of axis lengths $A_k(\Mel)$. The constant is determined so that the ellipsoid encloses $\Mel$. The coordinate system is redefined along the new axis lengths $A_k(\Mel)$, and the center is reset to the center-of-mass of the particles inside the new ellipsoid. Starting from the sphere centered on the center-of-mass of the FOF members, the above procedure is iterated until all the eigenvalues converge within one percent. In the above procedure, we use the all the particles {\it including substructures and non-FOF members}.

In literature, there are several methods to determine the axis lengths $A_k$ of simulated halos, including isodensity surfaces and other definitions of mass tensors \citep[JS02;][]{Despali14,Ludlow14,Bonamigo15,Vega16}. We decide to adopt the definition (\ref{Ixx}), and we explain why we prefer this estimator of the non-sphericity in what follows.

An alternative method to determine $A_k$ is the direct fitting to local isodensity surfaces, as adopted by JS02. Since the shape of isodensity surface is sensitive to substructures around halos, the removal of substructures is required in this method. The goal of this paper is, however, to construct the PDF of projected non-sphericity of halos for observational applications. Since it is difficult to definitely remove the effect of substructures in real observations, we do not use isodensity surfaces in the later sections.

We note, however, that the fitting to isodensity surfaces yields similar results to the mass tensor $I=\sum xx$ after substructures are removed. Figure \ref{mtcomp} shows the main halo of the same single-halo in top-left panel of Figure \ref{class} {\it without substructures}. We also plot the projections of the two ellipsoids with the same mass determined by the mass tensor $I=\sum xx$ (green) and the isodensity surface $\rho=100\rho_{\rm crit}$ (red), where $\rho_{\rm crit}$ is the cosmic critical density. The two ellipsoids are similar, indicating that the fitting to isodensity surfaces is an effective method to determine $A_k$ if substructures are removed.

Slightly different versions of mass tensors are also used in literature, including the following two;
\begin{equation}
\hat{I}_{\alpha\beta}=\sum^N_{i=1} \frac{x^{(i)}_{\alpha}x^{(i)}_{\beta}}{|\bm{x}^{(i)}|^2}\equiv\sum_i n^{(i)}_{\alpha}n^{(i)}_{\beta}
\label{Inn}
\end{equation}
and
\begin{equation}
\tilde{I}_{\alpha\beta}=\sum_i \frac{x^{(i)}_{\alpha}x^{(i)}_{\beta}}{[R_e^{(i)}]^2}\equiv\sum_i \tilde{n}^{(i)}_{\alpha}\tilde{n}^{(i)}_{\beta},
\label{Iee}
\end{equation}
where
\begin{equation}
R^{(i)}_e=\left[\left(\frac{x^{(i)}_1}{A_1}\right)^2+\left(\frac{x^{(i)}_2}{A_2}\right)^2+\left(\frac{x^{(i)}_3}{A_3}\right)^2\right]^{1/2}
\end{equation}
is the ellipsoidal distance of the $i$-th particle. 

To discuss the difference between the three mass tensors $I=\sum xx$, $\hat{I}=\sum nn$ and $\tilde{I}=\sum \tilde{n}\tilde{n}$, we consider a ``self-similar'' density distribution. Throughout this paper, we refer to the density distribution that is expressed by concentric ellipsoids with the same axis ratio and orientation as ``self-similar'' distribution.

For example, in the two-dimensional space, for a self-similar ellipse with axis lengths $p$ and $q$, the two-dimensional counterparts of the mass tensors $I=\sum xx$, $\hat{I}=\sum nn$ and $\tilde{I}=\sum\tilde{n}\tilde{n}$ yield ellipses with axis ratio $p/q$, $\sqrt{p/q}$ and $p/q$, respectively. Although $\hat{I}=\sum nn$ can be used as an estimator of the non-sphericity of halos, it does not reproduce the axis ratio of isodensity surfaces even for a self-similar density distribution.  In contrast, $\tilde{I}=\sum \tilde{n}\tilde{n}$ reproduces the isodensity surfaces of a self-similar density distribution. If the density distribution is not self-similar, however, the weighting by $R_e^{(i)}$ in Equation (\ref{Iee}) becomes inappropriate for inner regions. Hence we adopt the mass tensor $I=\sum xx$, which is free from such a weighting scheme and reproduces the isodensity surfaces of a self-similar density distribution.

The definition of mass tensor varies with authors in the previous literature, and therefore the applied method of triaxial modelling in each study should be carefully noticed. For example, JS02 fitted ellipsoids to isodensity surfaces of their simulated halos. Also, the mass tensors $I=\sum xx$, $\hat{I}=\sum nn$, $\tilde{I}=\sum\tilde{n}\tilde{n}$ are considered by \cite{Despali14}, \cite{Ludlow14} and \cite{Vega16}, respectively (although the results of \cite{Vega16} are mainly based on $I=\sum xx$). We emphasize that these results should not be {\it quantitatively} compared unless the same method of triaxial modelling is applied.

\section{Confrontation of EC Prediction against N-body Results}

\subsection{Ellipsoidal collapse model}

Before presenting the comparison between EC and the N-body results, we briefly summarize the basic framework of EC \citep{White79,Bond96}. For definiteness, we adopt the notation by \cite{Rossi11}.

EC describes the evolution of a homogeneous ellipsoid, embedded with a tidal field. The tidal field is characterized by the eigenvalues of the tensor $\nabla_{ij}\phi/(4\pi G\bar{\rho}a^3)$, where $\phi$, $\bar{\rho}$ and $a$ denote the gravitational potential, the mean matter density, and the scale factor, respectively. The differentiation by $\nabla_{ij}$ is operated in the comoving coordinate system. We denote the eigenvalues of the tensor by $\lambda_k$ ($k=1$, 2, 3; $\lambda_1\ge\lambda_2\ge\lambda_3$).

In the linear regime, the density contrast $\delta$ is given by $\sum_k\lambda_k$, and $\lambda_k$, $\delta$ and $\phi$ grow in proportion to the linear growth rate $D(t)$. Therefore, at the initial time $t_{\rm ini}$ where the linear regime holds, the axis lengths of the ellipsoid $A_k$ ($k=1$, 2, 3) satisfy the following equations:
\begin{equation}
A_k(t_{\rm ini})=a(t_{\rm ini})(1-\lambda_k(t_{\rm ini}))
\label{aini}
\end{equation}
and
\begin{equation}
\frac{dA_k(t_{\rm ini})}{dt}=H(t_{\rm ini})\left[A_k(t_{\rm ini})-a(t_{\rm ini})\lambda_k(t_{\rm ini})\left.\frac{d\ln D}{d\ln a}\right|_{t=t_{\rm ini}}\right],
\label{vini}
\end{equation}
where $H(t)$ is the Hubble parameter.

Then the axis lengths $A_k$ evolve according to the following equation of motion:
\begin{eqnarray}
\frac{d^2A_k(t)}{dt^2} & = & \Omega_{\Lambda,0} H_0^2A_k(t)\nonumber\\
& & -4\pi G\bar{\rho}(t)A_k(t)\left[\frac{1+\delta(t)}{3}+\frac{b'_k\delta(t)}{2}+\lambda'_{{\rm ext},k}(t)\right].
\label{eom}
\end{eqnarray}
The above equation of motion implies that the ellipsoid does not rotate with respect to the tidal field. Therefore the relation $A_3\ge A_2\ge A_1$ is conserved all the time since $\lambda_1\ge\lambda_2\ge\lambda_3$ at the initial time.

In the equation of motion (\ref{eom}), the interior tidal force $b'_k$ within the ellipsoids is computed by
\begin{equation}
b'_k(t)=\prod_j A_j(t)\int_0^\infty\frac{d\tau}{(A_k^2(t)+\tau)\prod_j \sqrt{A_j^2(t)+\tau}}-\frac{2}{3}.
\label{intide}
\end{equation}
Also, the exterior tidal force $\lambda'_{{\rm ext},k}$ is described by
\begin{equation}
\lambda'_{{\rm ext},k}(t)=\frac{D(t)}{D(t_{\rm ini})}\left[\lambda_k(t_{\rm ini})-\frac{\delta(t_{\rm ini})}{3}\right].
\label{extide}
\end{equation}
Equation (\ref{intide}) is the exact expression only for the homogeneous density as considered here. On the other hand, Equation (\ref{extide}) assumes the exact linear growth regime even when the later evolution may not be the case. Unlike $\lambda'_{{\rm ext},k}$, the density contrast $\delta(t)$ is calculated at each time so that the mass inside the ellipsoid $(4\pi\bar{\rho}/3) (1+\delta)A_1A_2A_3$ is constant. For the spherical case ($\lambda_1=\lambda_2=\lambda_3=\delta_{\rm ini}/3$ and $A_1=A_2=A_3=R$), both $b'_k$ and $\lambda'_{{\rm ext},k}$ vanish, and the equation of motion simply reduces to $d^2R/dt^2=\Omega_{\Lambda, 0} H_0^2R-(4\pi/3)G\bar{\rho}(1+\delta)R$.

According to Equation (\ref{eom}), all the axis lengths $A_k$ eventually collapse to zero, as in the spherical case. Therefore an additional assumption is needed to predict the eventual axis lengths $A_k$. In the spherical collapse model, it is conventionally assumed that the final (virial) radius $r_{\rm vir}$ and overdensity $\Delta_{\rm vir}$ of a homogeneous sphere are computed from the virial theorem.

In the case of EC, however, there may be no widely accepted treatment of anisotropic virialization of different axes. In this paper, we adopt the one proposed by \cite{Bond96}.  They assumed each $A_k$ {\it separately} stops collapsing when $A_k$ reaches $a(t)\times (\Delta_{\rm vir})^{-1/3}$, using the virial overdensity $\Delta_{\rm vir}(z=0)$ in the {\it spherical} virial theorem. Such an ellipsoid corresponds to a halo which is virialized at $z=0$. For the current set of cosmological parameters, $\Delta_{\rm vir}=355.4$ and so $(\Delta_{\rm vir})^{-1/3}=0.144$.

In summary, EC describes the evolution of a homogeneous and isolated ellipsoid, based on the liner growth of density fluctuations. The treatment of the virialization is based on the non-trivial assumption that each axis separately virialize; the axis lengths $A_k$ at low redshifts ($z\lesssim1$) are determined mainly by this virialization criterion. In the next subsection, we compare the evolution of the individual simulated halos with the EC prediction on the object-wise basis for the first time.

\subsection{Comparison of evolution of individual halos with EC prediction}

On the basis of the ellipsoids defined via the mass tensor $I=\sum xx$ (Equation (\ref{Ixx})), we compare the evolution of the individual simulated halos with the prediction of EC on the object-wise basis. For each FOF halo identified at $z=0$, we trace back the positions of the FOF member particles to each redshift. We then determine an ellipsoid of mass $\MFOF$ at each redshift via the mass tensor $I=\sum xx$ {\it by using all the particles including non-FOF particles}. We first choose the center of the calculation of $I=\sum xx$ as that of the FOF member particles of the current halo at the corresponding redshift, and then perform the iteration until it is converged. Throughout this paper, we call the ellipsoids determined at $z\neq0$ through the above procedure ``protohalos'' of each FOF halo. Note that the protohalos are {\it not} halos identified by the FOF algorithm at each redshift.

\begin{figure}[h]
\begin{center}
\subfigure{
\FigureFile(50mm,50mm){ecol8.eps}
\qquad
\FigureFile(50mm,50mm){rande8.eps}
}\\
\subfigure{
\FigureFile(50mm,50mm){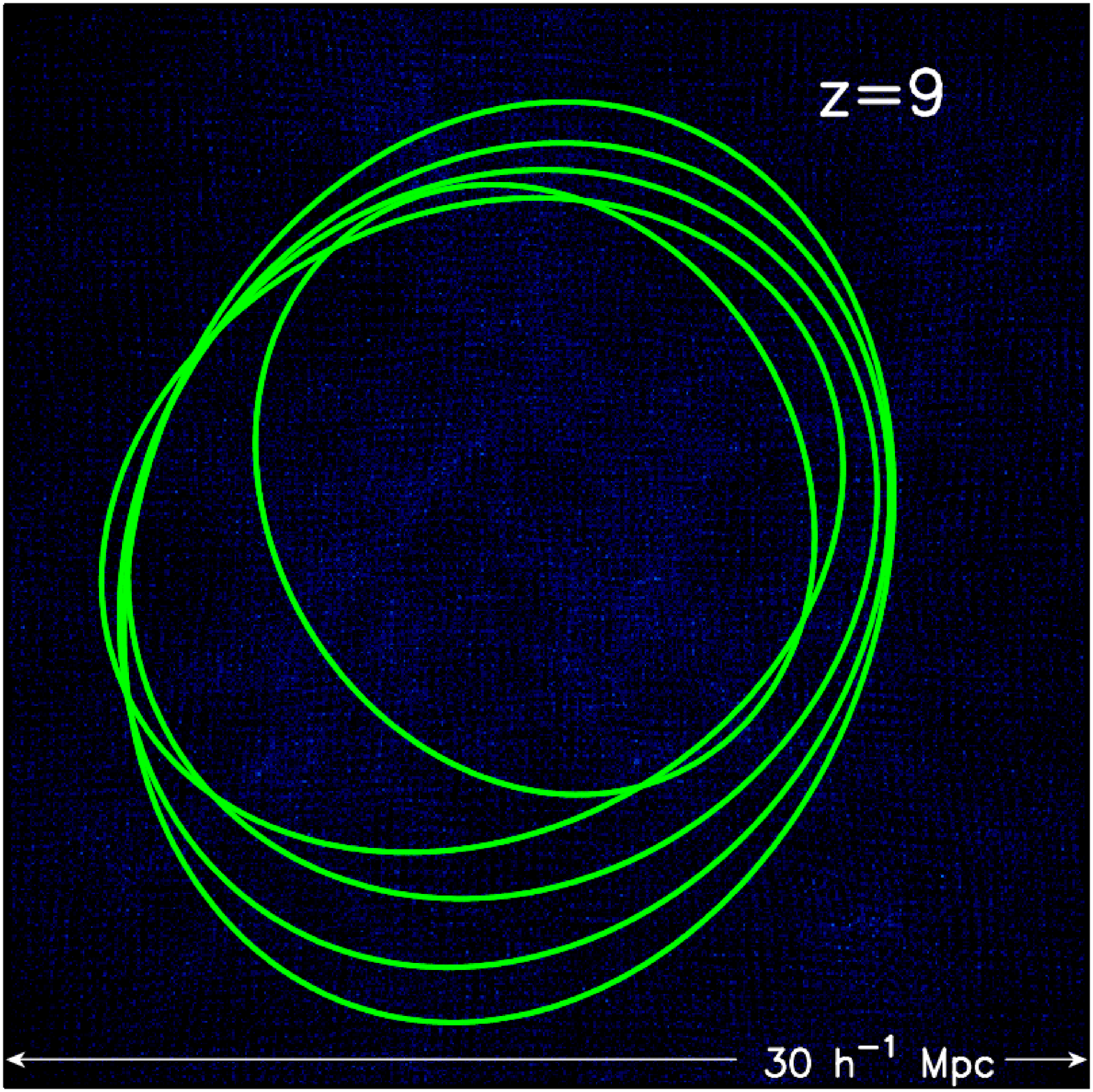}
\qquad
\FigureFile(50mm,50mm){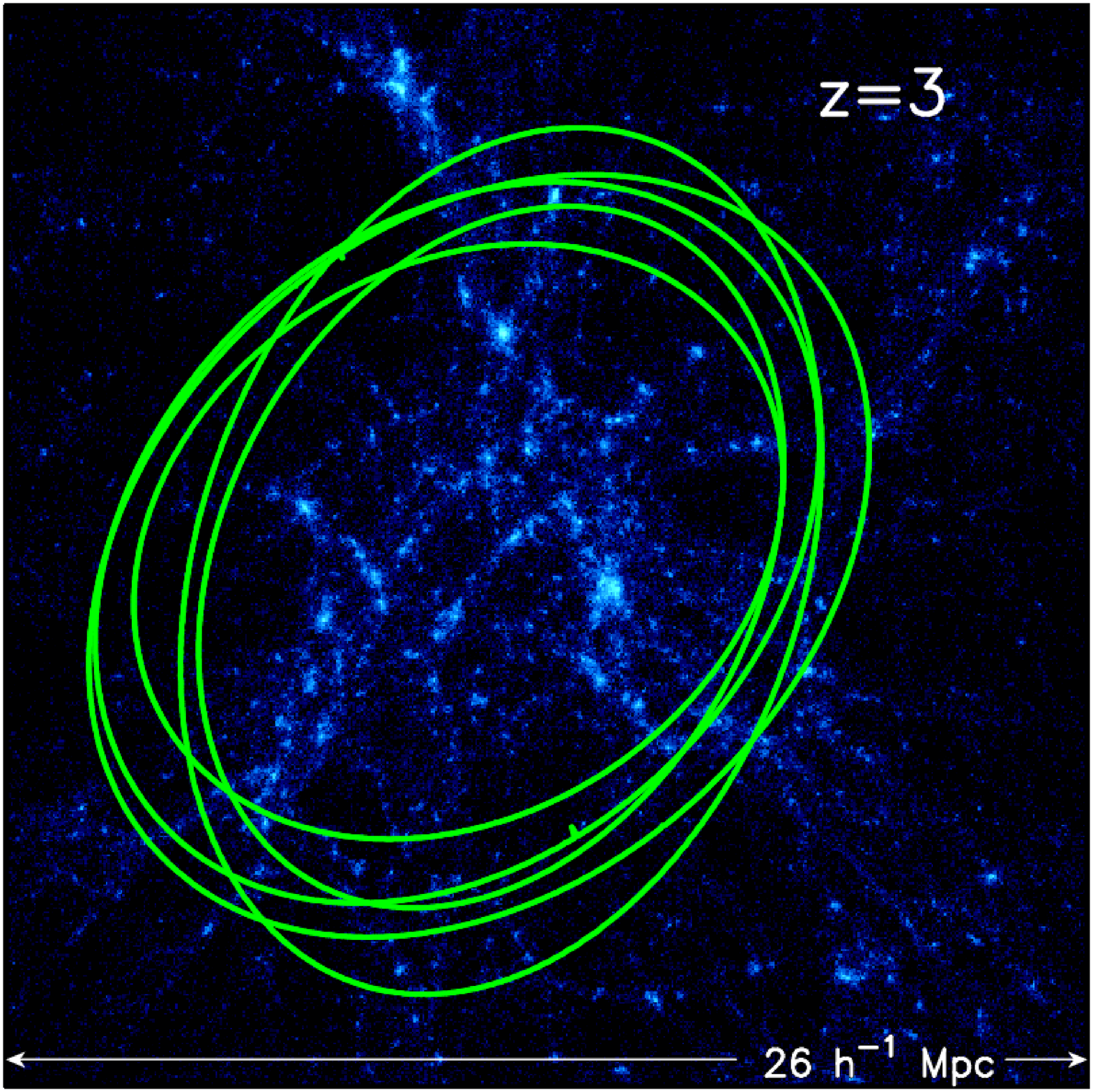}
}\\
\subfigure{
\FigureFile(50mm,50mm){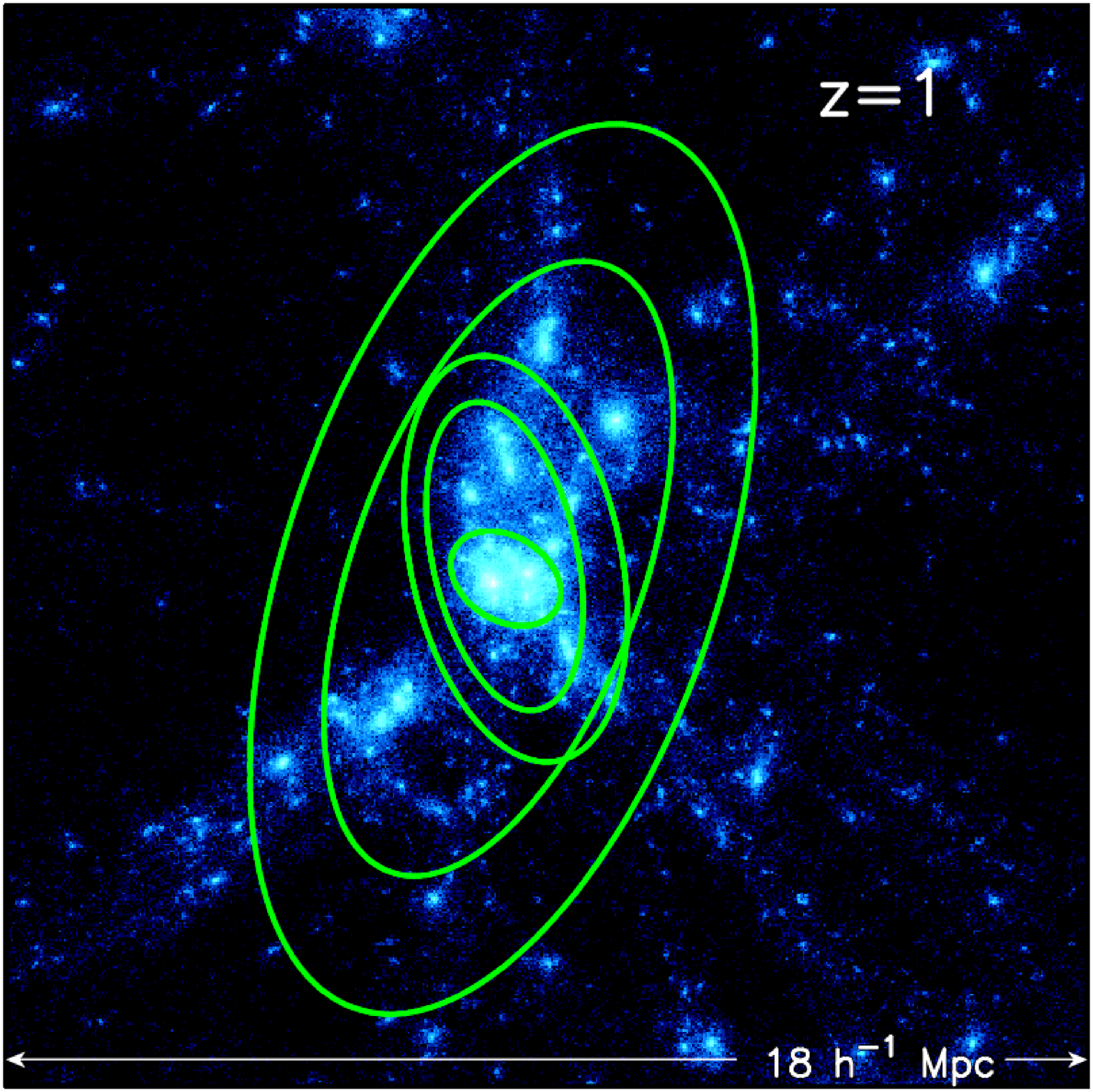}
\qquad
\FigureFile(50mm,50mm){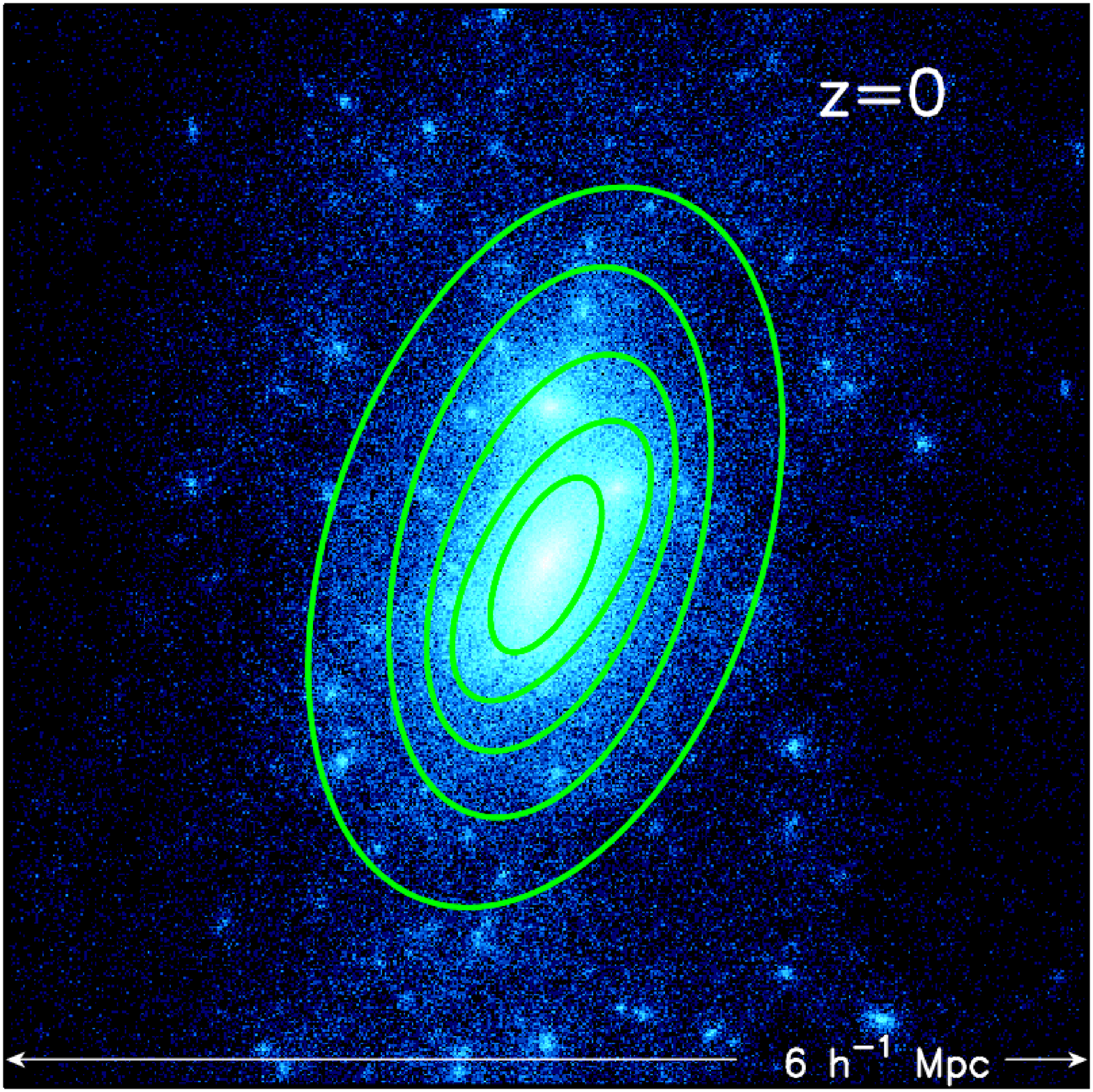}
}
\end{center}
\caption{Evolution of the single-halo
($\MFOF=8.43\times10^{14}h^{-1}M_\odot$) plotted in the top-left panel
of Figure \ref{class}. {\it top-left}: Evolution of the axis lengths
$A_k$. The squares indicate $A_k$ calculated from the mass tensor;
$A_1$, $A_2$, $A_3$ are colored in red, green, blue, respectively. The
solid lines indicate the EC prediction with the initial $\lambda_k$ are
calculated from $A_k$ at $z=99$ through Equation (\ref{aini}). The
dashed lines are also the EC prediction, but the initial $\lambda_k$ are
eigenvalues of the tensor $\nabla_{ij}\phi/(4\pi G\bar{\rho}a^3)$
calculated from the top-hat smoothed density field at the scale
$(3\MFOF/(4\pi\bar{\rho}))^{1/3}$. {\it top-right}: Evolution of the
axis ratio $A_1/A_3$ (cyan) and the ellipticity $e$ (magenta);
$A_1/A_3$: filled squares (simulation) and thin line (EC), $e$: open
circles (simulation) and thick line (EC). {\it middle and bottom}:
Density distributions around the halo at $z=9$, 3, 1 and 0. The
projections of the ellipsoids are determined by the mass tensor $I=\sum
xx$ for the five different mass scales inside the halos ($M=(s/5)
\MFOF$; $s=1$,..., 5) and plotted in green curves. All the particles
including the non-FOF members are shown, and used in determining the
ellipsoids.}  \label{ecol8}
\end{figure}

Figure \ref{ecol8} demonstrates the evolution of the single-halo in the top-left panel of Figure \ref{class} ($\MFOF=8.43\times10^{14}h^{-1}M_\odot$). The top-left panel shows the evolution of the axis lengths $A_k$ of the protohalos enclosing $\Mel=\MFOF$ in units of their initial values at $z=99$. The axis lengths $A_k$ determined by the mass tensor are plotted in filled squares; $A_1$, $A_2$ and $A_3$ are plotted in red, green and blue, respectively. The corresponding EC predictions are illustrated in solid and dotted lines with the same color as the simulation results. The solid lines adopt $\lambda_k$ evaluated from $A_k$ of the corresponding protohalo at $z=99$ through Equation (\ref{aini}); $\lambda_k=1-A_k(1-\delta_{\rm ini}/3)/(A_1A_2A_3)^{1/3}$. On the other hand, the dashed lines identify $\lambda_k$ with the eigenvalues of $\nabla_{ij}\phi/(4\pi G\bar{\rho}a^3)$ calculated from the top-hat smoothed density field at the scale $[3\MFOF/(4\pi \bar{\rho})]^{1/3}$, at the central position of the protohalo at $z=99$. The difference between the solid and dashed lines implies that the EC prediction is somewhat sensitive to the initial conditions, but the two sets of lines are roughly the same.

The simulation results and the EC prediction agree at least approximately for $z>9$. At around $z=9$, however, the simulation results begin to deviate from the EC prediction. As shown in the top-right panel, the corresponding ellipticity (magenta open circle) becomes larger than the EC prediction (magenta thick line), even though the linear regime still holds at $z\sim9$. The density distribution around the protohalo at $z=9$ is shown in the middle-left panel, and the projections of the ellipsoids with mass $\Mel/\MFOF=0.2$, 0.4, 0.6, 0.8, 1 are also plotted. The density distribution at $z=9$ is almost homogeneous, and so the triaxial modelling of the density distribution is not easy.

In EC, Equation (\ref{intide}) assumes the density distribution inside the ellipsoid is homogeneous. Inside the simulated halo, however, the density distribution becomes highly inhomogeneous from $z=3$ to $z=1$, as shown in the middle-right and bottom-left panels; particles falls into the central region of the protohalo along filamentary structures, and the innermost region ($\Mel\lesssim0.2\MFOF$) becomes highly denser. Due to the filamentary structures developed during these redshifts, the triaxial modelling is still a poor approximation of the density distribution. The inhomogeneity of density distribution is one of the reasons why the simulation results deviate from the EC prediction. In addition, the internal density distribution is far from self-similar; for example, the orientation of the inner ellipsoids at $z=1$ is considerably different from the outer ones.

Nevertheless, the evolution of the axis lengths $A_k$ very crudely follows the EC prediction up to the turn-around epoch ($z\sim1$) as seen in the top-left panel. Given that the various simplifications of EC, even this level of agreement between the simulation and EC may be surprising.

After the turn-around epoch, however, the simulation results more strongly deviate from the EC prediction. For example, the major axis $A_3$ (blue squares) rapidly increases and then decreases after $z=1$. Finally at $z=0$, the ellipticity $e$ is much larger than the EC prediction (top-right panel), although the triaxial modelling of the density distribution seems to work well at $z=0$ (bottom-right panel). The five ellipsoids at $z=0$ in the bottom-right panel are well aligned compared to $z=1$, but the density distribution is still not self-similar; the innermost ellipsoid is tilted with respect to the outermost one, and inner ellipsoids are slightly more elongated than outer ones.

As an another example, Figure \ref{ecol100} shows the results for another single-halo ($\MFOF=3.44\times10^{14}h^{-1}M_\odot$). Similarly to the case of Figure \ref{ecol8}, especially after the turn-around epoch, the simulation results substantially deviate from the EC prediction. As seen in the bottom and middle panels, the density distribution inside the halo is not self-similar, as well as the halo in Figure \ref{ecol8}.

We have found that the difference between the simulation and EC strongly depends on individual halos. Basically, however, the EC prediction very roughly reproduces the simulation results up to the turn-around. After that, the difference between the simulation and EC becomes larger.

One might expect that the difference between the simulation results and the EC prediction is larger for a multiple-halo than a single-halo. We have found that, however, this is not necessarily the case; the individuality of the halos is more noticeable.

Actually, the difference between the model prediction and the simulation results is not peculiar to EC. In \cite{Suto16}, we compare the evolution of the spherical radius of individual simulated halos with the prediction of the spherical collapse model. We then showed that the spherical collapse model fairly well reproduce the evolution of the simulation results up to the turn-around epoch. After the turn-around epoch, however, the evolution of simulated halos deviates from the prediction of the spherical collapse model. In this subsection, it has turned out that EC does not improve the difference between simulations and theoretical models. This rather implies that the spherical assumption works surprisingly well despite the highly non-spherical structure and evolution of halos.

In \cite{Suto16}, we also showed the difference is mainly caused by the velocity dispersion developed after the turn-around epoch. In order to better understand the difference between the simulation results and the EC prediction, we focus on the evolution of the simulated halos {\it after the turn-around epoch} ($z\sim 1$). Because the difference between the simulation results and the EC prediction varies appreciably from halo to halo, we {\it statistically} compare them in the next section.

We note that several authors previously obtained the mean axis ratio of ``protohalos'' $\langle A_1/A_3\rangle\sim0.7$ at $z\sim49$ \citep{Porciani02,Ludlow14,Despali14}, while our ``protohalos'' have $\langle A_1/A_3\rangle\sim1$ at $z>49$. This difference simply comes from the different definition of ``protohalos''; the above authors defined a protohalo {\it solely} from particles that are destined to be members of an FOF halo identified at lower redshifts. As stated in Section 2.3, we calculate the mass tensor by using all the particles ({\it not only the FOF members}) within a sphere around the mass centroid, and the information from the FOF algorithm is used only for choosing the initial position of the mass centroid.

The definition of protohalos in simulations is not necessarily unique, and can be different depending on the aim of each study. \cite{Ludlow14} noted that the axis ratio of  their protohalos (positions of FOF members at z>0) significantly deviates from that expected from Equation (\ref{aini}). Thus they decided to adopt the initial axis ratio directly measured from their simulated protohalos, instead of Equation (\ref{aini}), and to solve EC. By doing so, they found the better agreement between the EC model and simulations statistically. In contrast, we are interested in the individual evolution of each halo, and aim at identifying and clarifying when and how N-body simulations and the (standard) EC model deviate for the same initial condition. Hence we have defined our protohalos so that they satisfy the initial condition for $A_k$, i.e., Equation (\ref{aini}).

\section{Evolution and Radial Profile of Axis Ratio}

\subsection{Evolution and mass dependence of non-sphericity of halos}

One of the well-known discrepancies between EC and simulations is the mass dependence of ellipticity of halos at $z=0$. \cite{Rossi11} calculated EC for initial conditions described by the Gaussian random field and reported that more massive halos have smaller ellipticity in EC, while those in simulations have larger ellipticity at $z=0$.

We examine the evolution of axis ratio $A_1/A_3$ and ellipticity $e$ of our simulated halos. We have found that the initial $\lambda_k(z=99)$ measured from the simulation precisely reproduces the prediction for the Gaussian random field; more massive protohalos have smaller ellipticity at $z=99$. Also, more massive halos indeed have larger ellipticity on average at $z=0$ in our simulation, as reported in the previous studies \cite{Jing02,Despali14,Bonamigo15,Vega16}. Therefore the dependence of the non-sphericity of the simulated halos on their mass has changed sometime before the present time.

\begin{figure}[h]
\begin{center}
\subfigure{
\FigureFile(50mm,50mm){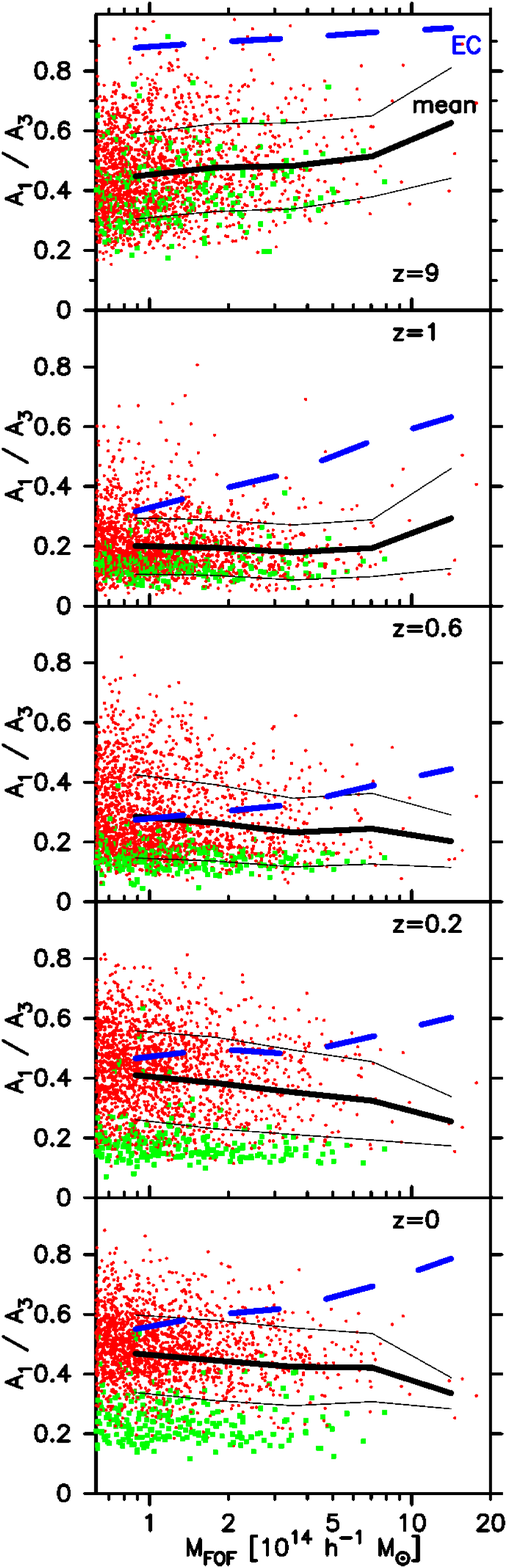}
\qquad
\FigureFile(50mm,50mm){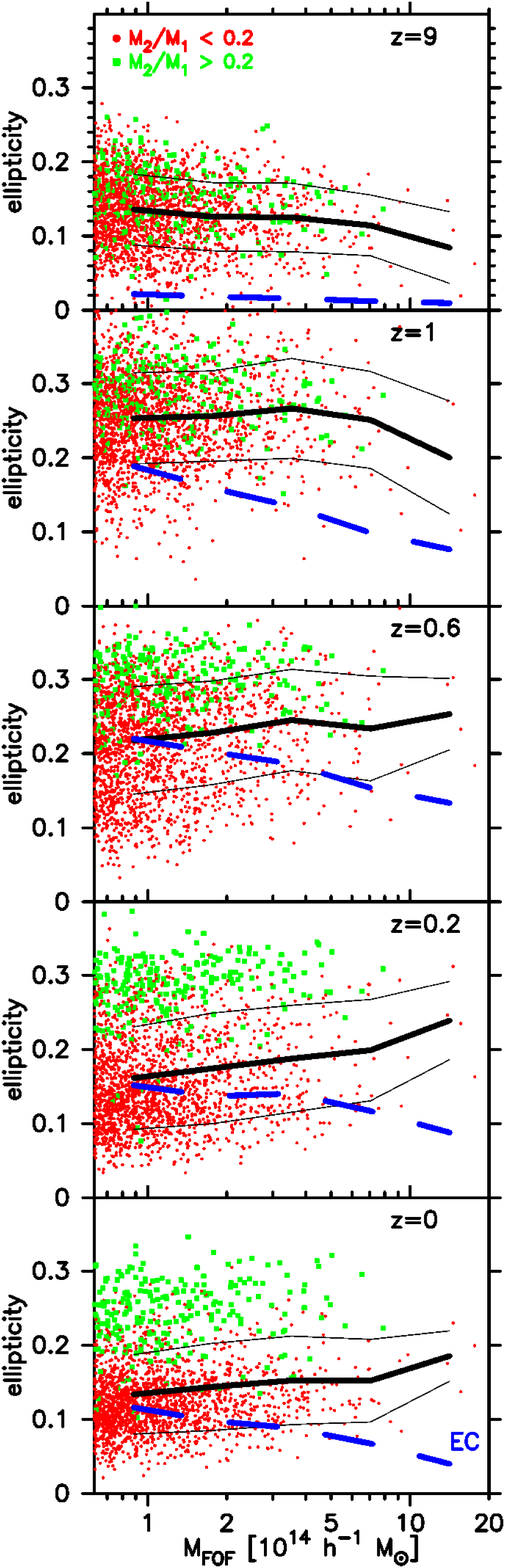}
}
\end{center}
\caption{Axis ratio $A_1/A_3$ (left) and ellipticity $e$ (right) of each
halo against its FOF mass $\MFOF$ at the five different redshift ($z=9$,
1, 0.6, 0.2, 0). Each symbol indicates the result for each of the 2004
simulated halos; red circle are single-halos ($M_2/M_1<0.2$), green
squares are for multiple-halos ($M_2/M_1>0.2$). The thick and thin solid
lines indicate the mean and the standard deviation, respectively, for
all the halos. For comparison, the blue dashed line indicates the EC
prediction, where the initial $\lambda_k$ are the eigenvalues of the tensor $\nabla_{ij}\phi/(4\pi G\bar{\rho}a^3)$ calculated from the top-hat smoothed density field at the scale $(3\MFOF/(4\pi\bar{\rho}))^{1/3}$. Note that the multiplicity ($M_2/M_1$) of the halos is determined only at $z=0$.}  
\label{eevol}
\end{figure}

Figure \ref{eevol} demonstrates the mass dependence of axis ratio $A_1/A_3$ and ellipticity $e$ at $z=9$, 1, 0.6, 0.2, 0. Each symbol indicates $A_1/A_3$ or $e$ of each halo; red circles are for single-halos ($M_2/M_1<0.2$) and green square are for multiple-halos ($M_2/M_1>0.2$). Note that, for $z\neq0$, the results are for the protohalos of each FOF halo identified at $z=0$.

The thick solid line illustrates the averaged value $\langle A_1/A_3\rangle$ or $\langle e\rangle$ over all the simulated halos with the root-mean-square scatter shown in thin lines. We have found that $\langle A_1/A_3\rangle$ and $\langle e\rangle$ only slightly change if we exclude the multiple-halos, although they are systematically less spherical than single-halos. This is because the fraction of the multiple-halos is small ($\sim10$\%). The blue dashed line indicates the EC prediction from the initial condition $\lambda_k$ calculated from $A_k$ at $z=99$ through Equation (\ref{aini}).

As shown in the top-left panel of Figure \ref{eevol}, at $z=9$, more massive halos have larger $A_1/A_3$ both in EC and the simulation results, reflecting the tendency at the initial time. The large scatter for the symbols implies the strong individuality of halos, i.e., the mass dependence of axis ratio is clear only when it is seen {\it statistically}. The simulation results have systematically smaller values of $A_1/A_3$ than the EC prediction, implying that the axis ratio $A_1/A_3$ of the majority of individual halos deviate from the EC prediction even at around $z=9$, as in the top-right panel of Figure \ref{ecol8}.

\begin{figure}[h]
\begin{center}
\subfigure{
\FigureFile(50mm,50mm){ecol100.eps}
\qquad
\FigureFile(50mm,50mm){rande100.eps}
}\\
\subfigure{
\FigureFile(50mm,50mm){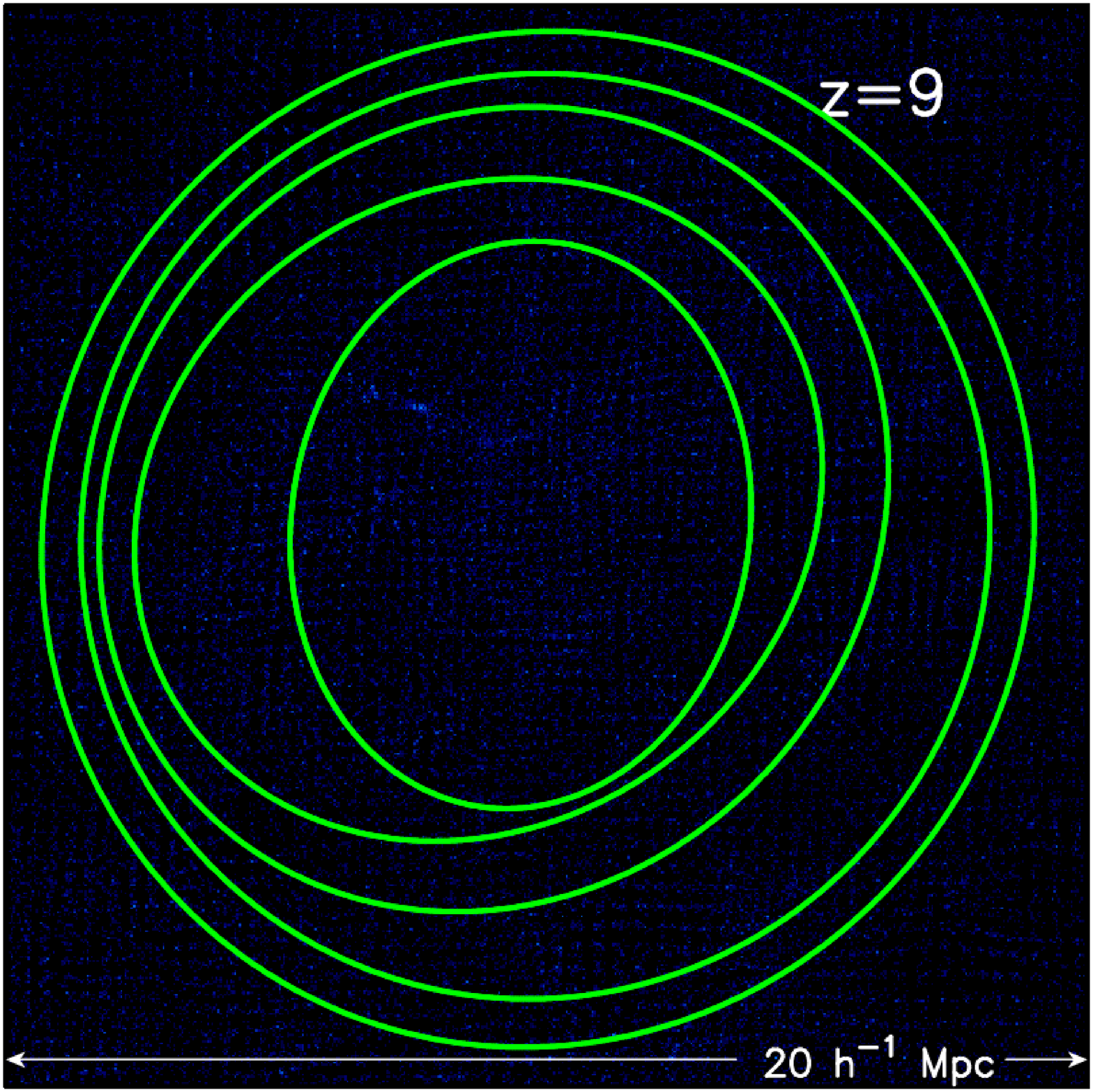}
\qquad
\FigureFile(50mm,50mm){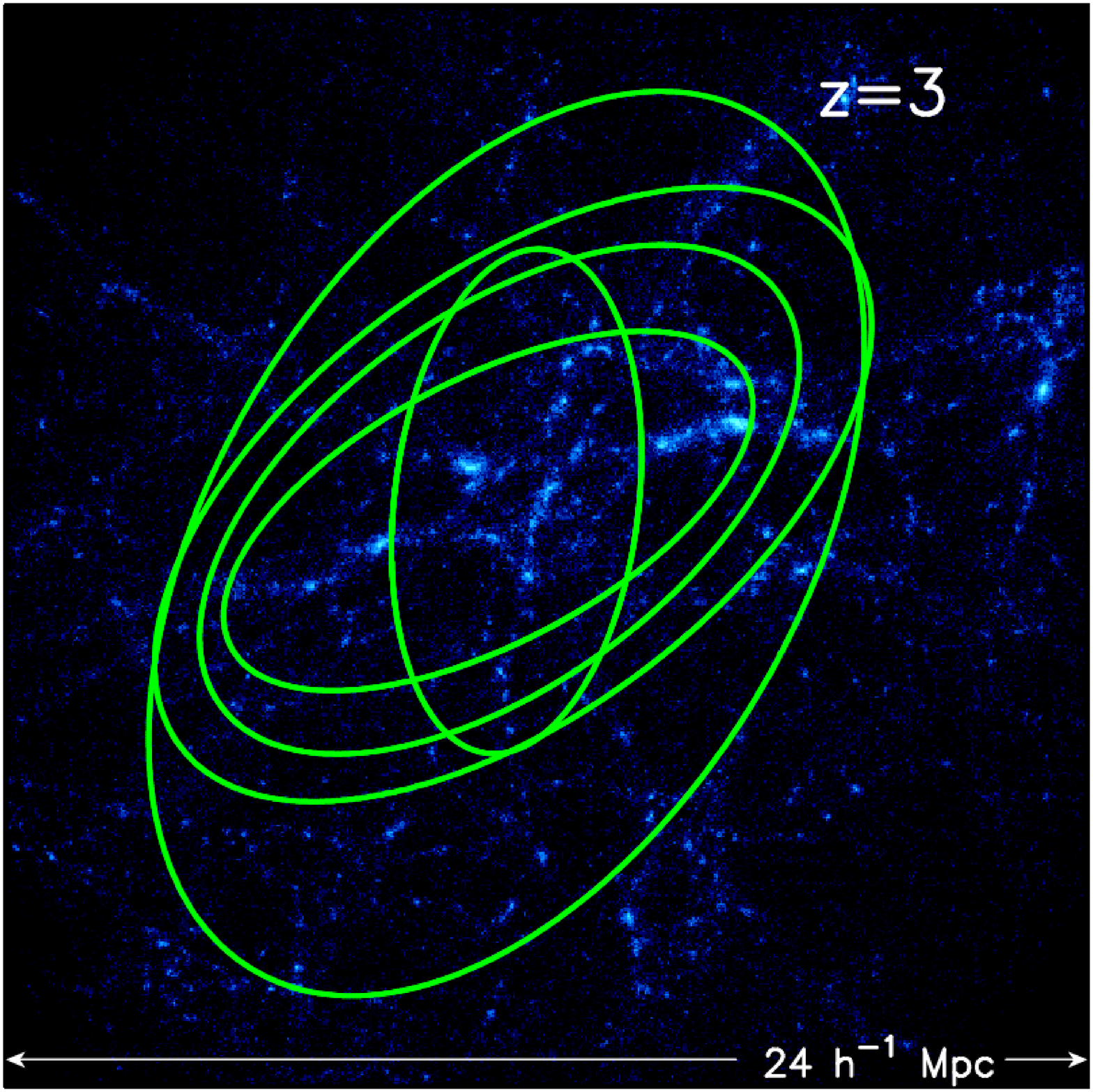}
}\\
\subfigure{
\FigureFile(50mm,50mm){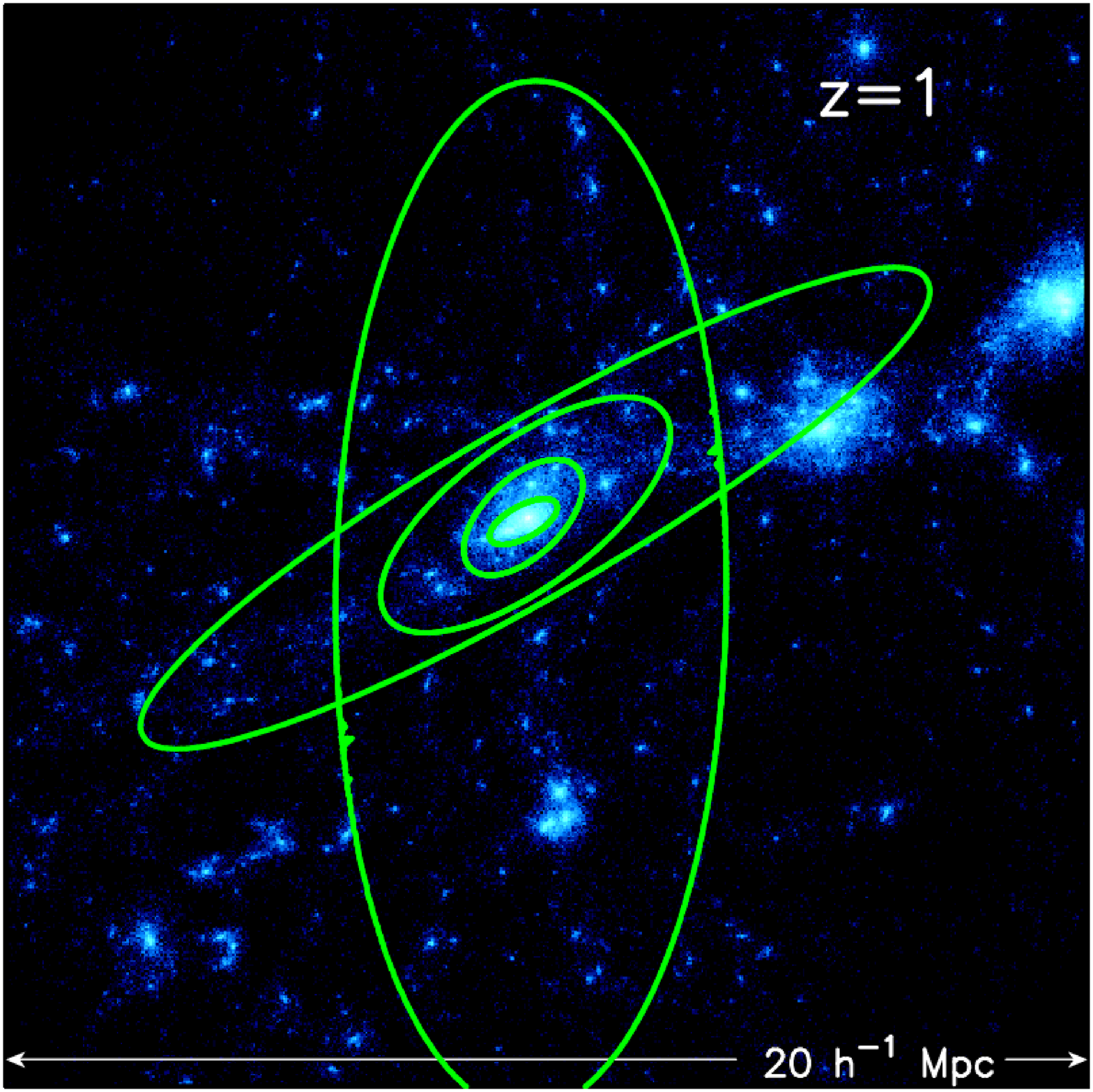}
\qquad
\FigureFile(50mm,50mm){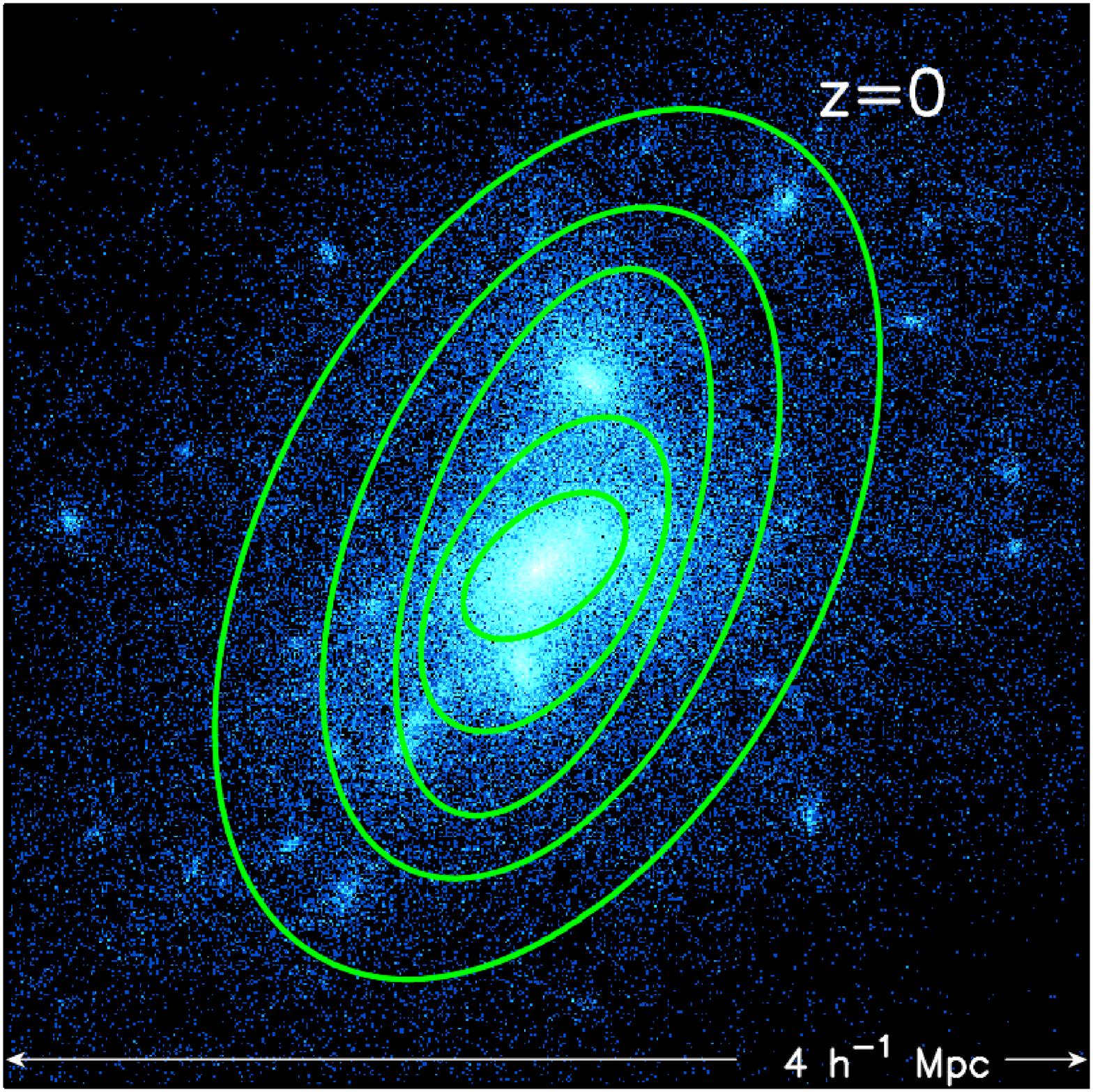}
}
\end{center}
\caption{Same as Figure \ref{ecol8}, but for another single-halo
($\MFOF=3.44\times10^{14}h^{-1}M_\odot$).}  \label{ecol100}
\end{figure}

At $z=1$ (second-top panel), the mass dependence is preserved in EC, but it becomes weaker at small mass scales for the simulated halos. From $z=0.6$ to $z=0$ (bottom three panels), $\langle A_1/A_3\rangle$ becomes gradually larger. The increase of $A_1/A_3$ is predicted by EC as in the top-right panels of Figures \ref{ecol8} and \ref{ecol100}, although the values of $A_1/A_3$ are much different between EC and the simulation.

Most importantly, the mass dependence of $\langle A_1/A_3\rangle$ of the simulated halos exhibits a clear {\it transition} after $z=1$; the mass dependence becomes even weaker, and finally at $z=0$, massive halos tend to be less spherical, opposite to that at the initial time. In contrast, the mass dependence of $\langle A_1/A_3\rangle$ in EC is preserved from the initial time to the present time; massive halos are more spherical. The mass dependence of $\langle e\rangle$ exhibits a similar transition to $\langle A_1/A_3\rangle$, as shown in the right panels of Figure \ref{eevol}. The redshift $z=1$ corresponds, on average, to the turn-around epoch where the difference between the EC prediction and the evolution of individual halos becomes large (see Figures \ref{ecol8} and \ref{ecol100}). We then expect that a similar transition of the mass dependence of $\langle A_1/A_3\rangle$ or $\langle e\rangle$ occurs earlier at inner mass scales of the halos, since inner regions turn-around earlier than outer regions.

\begin{figure}[h]
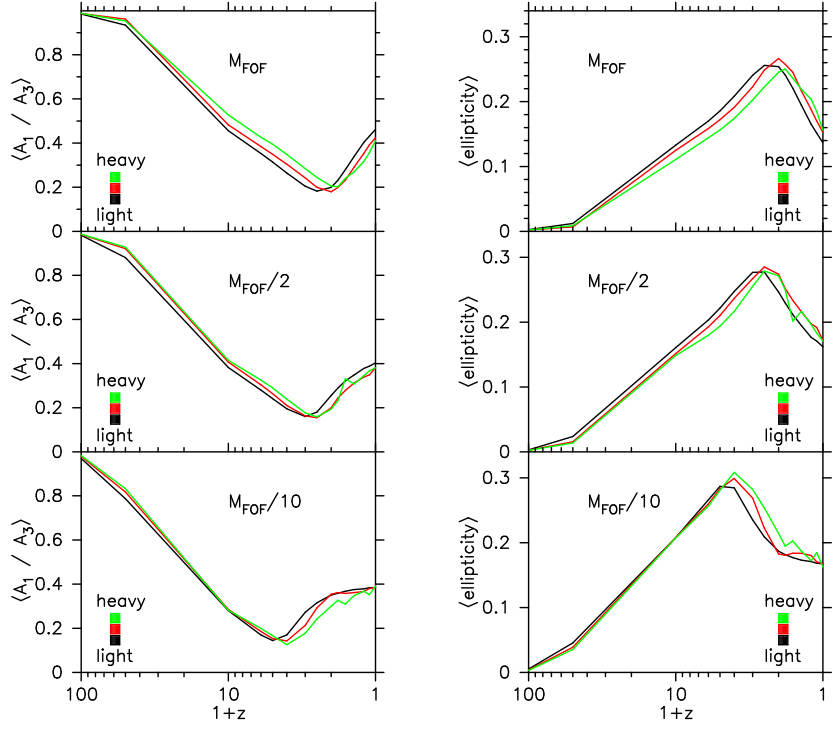

\begin{center}
\subfigure{
\FigureFile(50mm,50mm){mda1a3.eps}
\qquad
\FigureFile(50mm,50mm){mde.eps}
}
\end{center}
\caption{Evolution of the axis ratio $\langle A_1/A_3\rangle$ (left) and
ellipticity $\langle e\rangle$(right), averaged over the three different
mass ranges ($\MFOF>2.5\times10^{14}h^{-1}M_\odot$ ; green,
$1.25\times10^{14}h^{-1}M_\odot<\MFOF<2.5\times10^{14}h^{-1}M_\odot$;
red and
$6.25\times10^{13}h^{-1}M_\odot<\MFOF<1.25\times10^{14}h^{-1}M_\odot$;
black) at the three different mass sales ($\MFOF$; top, $\MFOF/2$;
middle, $\MFOF/10$; bottom).}
\label{evolmz}
\end{figure}

To confirm this, we compute the $A_k(\Mel)$ at the mass scales $\Mel(<\MFOF)$ for each halo. Figure \ref{evolmz} compares the evolution of $\langle A_1/A_3\rangle$ and $\langle e\rangle$ at the three different mass scales $\Mel=\MFOF$, $\MFOF/2$ and $\MFOF/10$. The values of $A_k$ are averaged over the three different mass ranges; heavy: $\MFOF>2.5\times10^{14}h^{-1}M_\odot$ (green), intermediate: $1.25\times10^{14}h^{-1}M_\odot<\MFOF<2.5\times10^{14}h^{-1}M_\odot$ (red) and light: $6.25\times10^{13}h^{-1}M_\odot<\MFOF<1.25\times10^{14}h^{-1}M_\odot$ (black).

The top-left panel of Figure \ref{evolmz} illustrates the redshift evolution of $\langle A_1/A_3\rangle$ at $\Mel=\MFOF$. At $z=99$, massive halos tend to be less spherical. Keeping this tendency, $\langle A_1/A_3\rangle$ decreases up to $z\sim1$, corresponding to the turn-around epoch. After that, $\langle A_1/A_3\rangle$ begins to increase and its mass dependence changes, as seen in Figure \ref{eevol}.

The middle-left and the bottom-left panels of Figure \ref{evolmz} show the results for the mass scales $\Mel=\MFOF/2$ (middle) and $\MFOF/10$ (bottom), respectively. As seen in Figure \ref{evolmz}, more massive halos become less spherical at $z\sim 1$. At $\Mel=\MFOF/2$ and $\Mel=\MFOF/10$, a similar change in the mass dependence of $\langle A_1/A_3\rangle$ occurs at $z\sim2$ and $z\sim4$, respectively. Indeed, the mass dependence of $\langle A_1/A_3\rangle$ changes earlier at inner mass scales. Similar things occur also in the mass dependence of $\langle e\rangle$, as shown in the right panels of Figure \ref{evolmz}.

The redshift where the mass dependence of $\langle A_1/A_3\rangle$ and $\langle e\rangle$ changes roughly corresponds to the turn-around epoch for each mass scale. Hence the change in the mass dependence of $\langle A_1/A_3\rangle$ is associated with the virialization process after the turn-around. We then suspect that the change in the mass dependence may be related to the development of the velocity dispersion after the turn-around epoch. Hence we examine the radial profile of the velocity dispersion after $z=1$ and compare it with the radial profiles of $\langle A_1/A_3\rangle$ and $\langle e\rangle$ in the next subsection.

\subsection{Radial profile of axis ratio inside FOF halos and the origin of the mass dependence of axis ratio}

\begin{figure}[h]
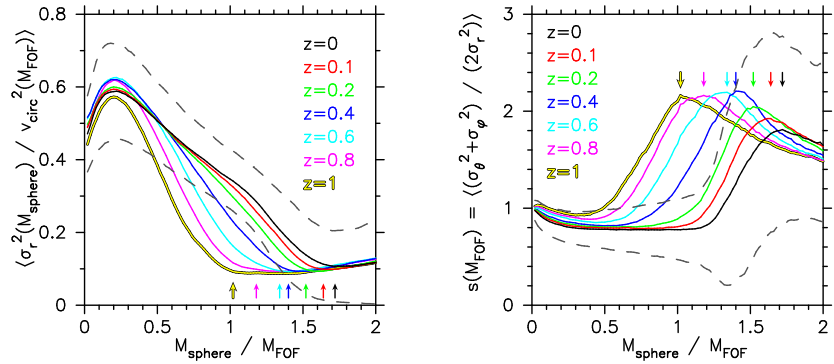

\begin{center}
\subfigure{
\FigureFile(50mm,50mm){sr.eps}
\qquad
\FigureFile(50mm,50mm){beta.eps}
}
\end{center}
\caption{Radial profiles of the radial velocity dispersion $\sigma_r^2$
(left) and the velocity isotropy measure
$s=(\sigma_\theta^2+\sigma_\varphi^2)/(2\sigma_r^2)$ (right), averaged
over the 2004 simulated halos, at the seven different redshifts; $z=1$,
0.8, 0.6, 0.4, 0.2, 0.1, 0. The velocity dispersion in the left panel is
normalized by the circular velocity $v_{\rm circ}^2(\MFOF)=G\MFOF/R_{\rm
FOF}$ of each halo at each redshift. The dashed lines indicate the
standard deviation for $z=0$. At each redshift, the mass scale where $s$
reaches a maximum is indicated by an arrow in both panels.}
\label{sprof}
\end{figure}

Figure \ref{sprof} shows the radial profiles of the radial velocity dispersion $\sigma_r^2$ and the ``velocity isotropy measure'' defined by $(\sigma_\theta^2+\sigma^2_\varphi)/(2\sigma_r^2)$ at $z=1$, 0.8, 0.6, 0.4, 0.2, 0.1, 0 (after the turn-around epoch). We here use spherical mass coordinate $M_{\rm sphere}$ for simplicity, and calculate each component of velocity dispersion in the spherical coordinate. In the left panel, the radial velocity dispersion $\sigma_r^2$ is normalized by the circular velocity $v_{\rm circ}^2(\MFOF)$ at $M_{\rm sphere}=\MFOF$:
\begin{equation}
v^2_{\rm circ}(\MFOF)=\frac{G\MFOF}{R_{\rm FOF}},
\end{equation}
where $R_{\rm FOF}$ is the radius of the sphere enclosing the mass $\MFOF$. We first compute the radial profiles of $\sigma_r^2/v^2_{\rm circ}$ and $(\sigma_\theta^2+\sigma^2_\varphi)/(2\sigma_r^2)$ for an individual halo, and then average further the radial profiles over the 2004 halos to obtain the ``mean'' radial profiles, $\langle\sigma_r^2/v^2_{\rm circ}\rangle$ and $s\equiv\langle(\sigma_\theta^2+\sigma^2_\varphi)/(2\sigma_r^2) \rangle$. Note that, for $z\neq0$, the results are for the protohalos of each FOF halo identified at $z=0$. We have confirmed that the radial profiles in Figure \ref{sprof} are almost unchanged even if we include/exclude the multiple-halos.

The left panel of Figure \ref{sprof} indicates that the radial velocity dispersion $\langle\sigma_r^2/v^2_{\rm circ}\rangle$ is larger at the innermost regions at every redshift. At around $M_{\rm sphere}=\MFOF$, $\sigma_r^2$ rapidly decreases and becomes roughly constant at outer regions. The small values of $\langle\sigma_r^2/v^2_{\rm circ}\rangle$ at outer regions can be attributed to the particles coherently falling toward the central region with small radial velocity dispersion.

In the right panel of Figure \ref{sprof}, the averaged radial profile of the velocity isotropy measure $s$ at each redshift has roughly three different regions. At the innermost region, $s$ is approximately unity, indicating that the velocity is almost isotropic. At around $M_{\rm sphere}=\MFOF$, $s$ rapidly increases, corresponding to the decrease of $\langle\sigma_r^2/v^2_{\rm circ}\rangle$ in the left panel. Then $s$ reaches a maximum. We indicate the maximum point by an arrow in the figure. Outside the maximum point, $s$ slowly decreases.

We indicate the location where the velocity isotropy measure $s$ reaches the maximum by an arrow also in the left panel. At this location, the radial profile of $\langle\sigma_r^2/v^2_{\rm circ}\rangle$ becomes roughly flat. We find that this location very roughly corresponds to the ``splash-back radius'' $r_{\rm sb}$ \citep{Adhikari14,Diemer14,More15}, although their agreement strongly depends on individual halos. We note that $r_{\rm sb}$ moves outward with time, indicating that the velocity dispersion develops and extends outward. We next examine how the radial profiles of the axis ratio $\langle A_1/A_3\rangle$ and $\langle e\rangle$ behaves inside and outside $r_{\rm sb}$.

\begin{figure}[h]
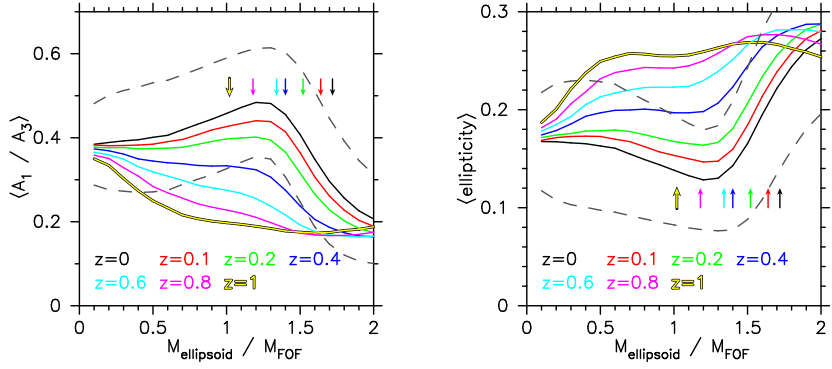

\begin{center}
\subfigure{
\FigureFile(50mm,50mm){a1a3prof.eps}
\qquad
\FigureFile(50mm,50mm){eprof.eps}
}
\end{center}
\caption{Radial profiles of the axis ratio $\langle A_1/A_3\rangle$
(left) and the ellipticity $\langle e\rangle$ (right), averaged over the
2004 simulated halos, at the seven different redshifts; $z=1$, 0.8, 0.6,
0.4, 0.2, 0.1, 0. The dashed lines indicate the standard deviation for
$z=0$. At each redshift, the {\it spherical} mass scale where $s$
reaches a maximum (Figure \ref{sprof}) is indicated by an arrow for both
panels.}  \label{eprof}
\end{figure}

Figure \ref{eprof} illustrates the radial profiles of axis ratio $\langle A_1/A_3\rangle$ and ellipticity $\langle e\rangle$ averaged over our 2004 halos for $z=1$, 0.8, 0.6, 0.4, 0.2, 0.1, 0. The horizontal axis $\Mel$ indicates the mass of ellipsoids determined by the mass tensor $I=\sum xx$ using internal and external density distributions for each halo. We refer to the sequence of $\langle A_1/A_3\rangle$ or $\langle e\rangle$ of such ellipsoids as ``radial profiles''. Note that the central position differs from inner to outer ellipsoids belonging to the same FOF halo (see bottom panels of Figures \ref{ecol8} and \ref{ecol100}). We have confirmed that the radial profiles in Figure \ref{eprof} are almost unchanged even if we include/exclude the multiple-halos.

The left panel of Figure \ref{eprof} shows the evolution of the radial profile of $\langle A_1/A_3\rangle$. At least after $z\sim 0.4$, the radial profile of $\langle A_1/A_3\rangle$ rapidly decreases beyond a certain mass scale around $\Mel\sim\MFOF$. Similarly, as shown in the right panel of Figure \ref{eprof}, the profile of ellipticity $\langle e\rangle$ rapidly increases there. This corresponds to the development of filamentary structures surrounding the halos (cf. the bottom-left panel of Figure \ref{ecol8}). The characteristic mass scale moves outward with time, and eventually becomes larger than $\MFOF$ after $z\lesssim0.4$.

We indicate the location where the velocity isotropy measure $s$ reaches a maximum, roughly corresponding to the splash-back radius $r_{\rm sb}$, at each redshift by an arrow in both panels of Figure \ref{eprof}. The characteristic mass scale in the radial profile of $\langle A_1/A_3\rangle$ or $\langle e\rangle$ roughly corresponds to $r_{\rm sb}$, given that $M_{\rm sphere}$ is not exactly identical to $\Mel$. These two mass scales may give a rough indication of the physical boundary of halos inside which the velocity dispersion has been developed.

Figures \ref{eevol} to \ref{eprof} imply that the mass dependence of axis ratio $\langle A_1/A_3\rangle$ changes almost simultaneously the velocity dispersion $\langle\sigma_r^2/v^2_{\rm circ}\rangle$ becomes larger. We note, however, that the halos have a significant mean ellipticity $\langle e\rangle$ inside the splash-back radius $r_{\rm sb}$. This may seem inconsistent with the fact that the velocity dispersion is almost isotropic at the innermost region (Figure \ref{sprof}). Hence some unknown mechanism other than the velocity anisotropy is needed to maintain the highly non-spherical density distribution of the halos, which remains as a puzzle.

We have adopted a simple version of EC. In contrast, several authors have proposed to refine EC, for instance, by a better mapping between the current halos and their corresponding ancestors at earlier epochs \citep{Borzyszkowski14,Ludlow14}, and by using a different empirical treatment of the virialization \citep{Angrick10}. While those approaches improve the agreement between the EC predictions and simulation results to some extent, they require additional ad-hoc assumptions, and are not necessarily satisfactory.  Thus we do not consider the modification of the EC model, but conclude  that the predictions of EC are not so robust, and that quantitative predictions of non-sphericities of clusters in precision cosmology era should be made via numerical simulations.

Hence we move on to detailed analyses of simulation results in the next section. Especially, we pay a special attention to the radial dependence of $\langle A_1/A_3\rangle$ and $\langle e\rangle$ in Figure \ref{eprof}. While inner regions are more spherical at $z=1$, inner regions are less spherical at $z=0$. This radial dependence may seem small, but indicates that the halos are not self-similar. We examine how the probability distribution function of $A_1/A_3$ depends on $\Mel$ in the next section.

\section{Probability Distribution Function of Axis Ratio}

\subsection{Minor-to-major axis ratio of triaxial ellipsoids}

Previously, JS02 measured the minor-to-major axis ratio $A_1/A_3$ of the isodensity surface at $\rho=2500 \rho_c$ (approximately corresponding to 0.3 $r_{\rm vir}$) of their simulated halos. They further assumed the self-similarity of the density distribution inside the halos, and obtained the following fitting formula at the virial mass $\Mvir$:
\begin{eqnarray}
P(A_1/A_3; \Mvir,z)=
\frac{1}{\sqrt{2\pi}0.113}\left(\frac{\Mvir}{M_*}\right)^{0.07\Omega_m(z)^{0.7}}\nonumber \\
\times\exp\left[\frac{[(A_1/A_3)(\Mvir/M_*)^{0.07\Omega_m(z)^{0.7}}-0.54]^2}{2(0.113)^2}\right],
\label{JS}
\end{eqnarray}
where $M_*(z)$ is the characteristic non-linear mass scale. The scale is determined so that the top-hat smoothed mass fluctuation $\sigma(M_*,z)$ becomes $\delta_c=1.68$, where $\delta_c$ is the linearly-extrapolated critical density contrast in the spherical collapse model.

In the previous section, however, we have seen that the axis ratio $A_1/A_3(\Mel)$ is not constant as a function of $\Mel$ (Figure \ref{eprof}). This result implies that the formula (\ref{JS}) may not be reliable since it is based on the self-similarity assumption. Therefore, we here quantitatively show the extent to which the departure from self-similarity affects the probability distribution function (PDF) of $A_1/A_3$.

Figure \ref{a1a3hist} illustrates the PDFs of axis ratio $A_1/A_3$ of our 2004 halos, determined by $I=\sum xx$ at the three different mass scales $\Mel=\MFOF$, $\MFOF/2$ and $\MFOF/10$. The simulated halos are classified into three categories according to the mass of the most massive substructure $M_2$ compared to that of the main halo $M_1$; $M_2/M_1<0.1$ (red), $0.1<M_2/M_1<0.2$ (blue), $M_2/M_1>0.2$ (green). The red and blue portions correspond to the single-halos defined in Section 2.2, and the green portion corresponds to the multiple-halos.

The bottom panel of Figure \ref{a1a3hist} shows the result for $\Mel=\MFOF/10$, approximately corresponding to the region enclosed by the isodensity surface $\rho=2500 \rho_c$. The PDF of our halos is shifted to the left compared to Equation (\ref{JS}). Their difference may be partly explained by the different methods of triaxial modelling of halos; the mass tensor and the isodensity surfaces.

As shown in the middle panel of Figure \ref{a1a3hist}, the PDF of $A_1/A_3$ for $\Mel=\MFOF/2$ is shifted to the right compared with that of $\MFOF/10$. Hence the region at $\Mel=\MFOF/2$ inside the halos is, on average, more spherical than $\MFOF/10$, corresponding to the radial profiles of $A_1/A_3$ and $e$ in Figure \ref{eprof}. Similarly, as shown in the top panel, the region $\Mel=\MFOF$ is even more spherical, clearly indicating that the PDF of $A_1/A_3$ depends on $\Mel$ due to the non-self-similarity of halos. Quantitatively, the mean value $\langle A_1/A_3\rangle$ at $\Mel=\MFOF/2$ and $\Mel=\MFOF/10$ is smaller by $\sim10$\% and $\sim15$\% than that at $\Mel=\MFOF$.

For $\Mel=\MFOF$, the PDF of our halos is similar to Equation (\ref{JS}), except for the fraction by the multiple-halos. This is most likely just a coincidence; the difference in $A_k$ by the mass tensor and the isodensity surfaces, and the radial profile of $A_1/A_3$ are accidentally compensated.  At $\Mel=\MFOF/2$ and $\MFOF$, the multiple-halos are significantly less spherical than single-halos. In contrast, the multiple-halos do not have such a tendency at $\Mel=\MFOF/10$. This is because the multiplicity of halos is determined by the amount of substructures with all the FOF members; For example, the region of $\MFOF/10$ of a halo comprising two comparable mass objects may include only one of them.

Previously, \cite{Vega16} calculated the PDF of $A_1/A_3$ of their simulated halos at the two different mass scales $\Mel\approx\MFOF$ and $\Mel\approx\MFOF/2$ {\it without} the self-similarity assumption. They then found that $\langle A_1/A_3\rangle$ at the inner mass scale is smaller by $\sim10$ \% than that at the outer mass scale. This is consistent with our results, although their methods of triaxial modelling and halo identification are slightly different from ours.

In observations, since the density distribution of halos is projected on the sky, the PDF of projected axis ratio is more relevant in interpreting observational data. In fact, OLS03 calculated the PDF of projected axis ratio by integrating imaginary halos whose axis lengths $A_k$ follow the formula (\ref{JS}). The scale dependence of the PDF of $A_1/A_3$ indicates that the self-similar assumption is not valid when constructing a PDF of projected axis ratio through a PDF of $A_1/A_3$, as employed previously \citep[e.g., OLS03;][]{Kawahara10}. Therefore, we instead directly measure the projected axis lengths by projecting the density distribution of the simulated halos in the next subsection.

\subsection{Axis ratio from projected density distribution} 

\begin{figure}[h]
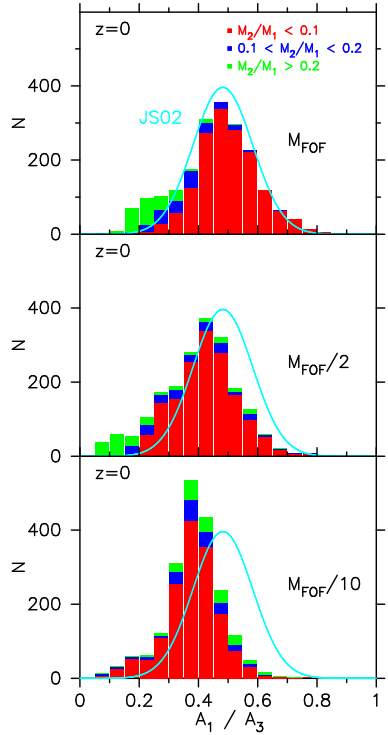

\begin{center}
\subfigure{
\FigureFile(50mm,50mm){a1a3hist.eps}
}
\end{center}
\caption{PDF of the minor-to-major axis ratio $A_1/A_3$ of triaxial
ellipsoid at $z=0$ for the three different mass scales: $\Mel=\MFOF$,
$\MFOF/2$ and $\MFOF/10$.  The histogram is divided by three types of
halos; $M_2/M_1<0.1$ (red), $0.1<M_2/M_1<0.2$ (blue), $M_2/M_1>0.2$
(green). The cyan curve shows the fitting formula of JS02 (Equation
(\ref{JS})) that is based on the isodensity surface $\rho=2500 \rho_c$,
approximately corresponding to $0.3 r_{\rm vir}$ and $\MFOF/10$.}
\label{a1a3hist}
\end{figure}

Figure \ref{twod} (a) shows the histograms of the projected axis ratio of our halos at $z=0$. Instead of $\MFOF$, $\MFOF/2$, $\MFOF/10$ in Figure \ref{a1a3hist}, we measure three observationally more relevant mass scales; $\Mvir$, $M_{500}$ and $M_{2500}$. The virial mass $\Mvir$ is defined as the mass of the sphere within which the averaged overdensity becomes $\Delta_{\rm vir} (z=0)=355.4$ times cosmic mean matter density, and $M_{500}$ and $M_{2500}$ are the masses of the sphere within which the mean density is 500 and 2500 times the cosmic critical density. Actual lensing halos are observed roughly up to the scale of $M_{500}$. Typically, $M_{500}\sim0.5\Mvir$ and $M_{2500}\sim0.2\Mvir$. In reality, these mass scales are measured from the projected density distribution on the sky, but here we determine them in the three-dimensional space for simplicity.

We determine an ellipse by using the two dimensional counterpart of $I=\sum xx$ from the projected density distributions of each halo along the $x$-, $y$- and $z$-axes of our simulation. We choose a rectangular box with the depth only along the line-of sight confined so that the region barely encloses all the FOF member particles. Therefore we consider all the particles in the box, but neglect the contribution from foreground and background particles outside the box. The particle number $N$ in Equation (\ref{Ixx}) is set so that the (projected) mass inside the ellipse becomes any of $\Mvir$, $M_{500}$ and $M_{2500}$. We call the axis lengths of the resulting ellipse $a_1$ and $a_2$ ($a_1<a_2$). Note that we obtain three values of $a_1/a_2$ for each halo.

The top panel of Figure \ref{twod} (a) shows the histogram for $\Mvir$ ($\approx \MFOF$) at $z=0$. The histogram is separately colored by single-halos ($M_2/M_1<$0.2) and multiple-halos ($M_2/M_1>$0.2). Due to the projection effect, the overall shape of the histogram is broader and more shifted to the right (rounder) than that of $A_1/A_3$ for $\MFOF$ in Figure \ref{a1a3hist}. Also, compared to $A_1/A_3$, the portion of the multiple-halos in the PDF is extended to the right; if two major components of a multiple-halo are along the line-of-sight, it may be regarded a single object from an observer.

We find that the histogram of projected axis ratio is well approximated by the beta distribution:
\begin{equation}
P(x; a,b)=\frac{x^{a-1}(1-x)^{b-1}}{B(a,b)},
\label{betad}
\end{equation}
where 
\begin{equation}
B(a,b)=\int_0^1x^{a-1}(1-x)^{b-1}dx
\end{equation}
is the beta function and $a$ and $b$ are parameters. The mean $\mu$ and the variance $\sigma^2$ of the beta distribution are given by
\begin{equation}
\mu=\frac{a}{a+b},\qquad\sigma^2=\frac{ab}{(a+b)^2(a+b+1)},
\end{equation}
respectively. Table 2 lists the parameters $a$ and $b$ along with the mean value and the standard deviation calculated from $a$ and $b$. The values of mean $\mu$ and variance $\sigma^2$ do not change much even when we include/exclude the multiple-halos. Such weak dependence on the multiplicity of halos is useful when the fitting formula is compared with real observational data, since it is difficult to determine the multiplicity of real halos, and to remove substructures in observations.

Our result should be compared with the PDF of $a_1/a_2$ by OLS03 that integrates the PDF of $A_1/A_3$ by JS02. We emphasize that the PDF of OLS03 is sensitive to the self-similarity assumption by JS02. When calculating the PDF of OLS03, we substitute $\Mvir=2\times 10^{14}h^{-1}M_\odot$ in Equation (\ref{JS}), corresponding to the mean mass of our sample. The PDF of OLS03 is plotted in blue dashed curve in Figure \ref{twod}. Since the PDF of $A_1/A_3$ for $\MFOF(\approx \Mvir)$ well follows the model of JS02 by coincidence (the top-panel of Figure \ref{a1a3hist}), the difference between OLS03 and the histogram is mainly due to the self-similarity assumption for the density distribution inside halos, adopted by JS02 and OLS03. This difference clearly demonstrates the importance of the projection effect.

The middle and bottom panels of Figure \ref{twod} (a) show the histograms for $M_{500}$ and $M_{2500}$, respectively, compared with OLS03. Since OLS03 assumes the self-similarity of density distribution, the blue-dashed curves in the three panels of Figure \ref{twod} (a) are the same. These histograms show that the inner region is slightly less spherical than the outer region. This dependence is similar to the case of $A_1/A_3$ (Figure \ref{a1a3hist}), but significantly weaker due to the projection.

The PDF of $a_1/a_2$ at $M_{500}$ and $M_{2500}$ are also well approximated by the beta distribution, and the best-fit parameters are listed in Table 2. It may seem that, for $M_{2500}$, the PDF of OLS03 is in better agreement in the simulation results.  Given that the significant difference in the bottom panel of Figure \ref{a1a3hist} at $\MFOF/10$, however, this is also just a coincidence, and rather implies the importance of the projection effect for non-self-similar halos.

We repeat the same analysis for $z=0.2$, 0.4 and 1. In doing so, we find halos by the FOF algorithm {\it at each redshift separately}; in the preceding sections, we have traced back the evolution of protohalos of each FOF halo identified at $z=0$, which does not correspond to real observational situations because observed halos are defined at each $z$. The multiplicity of the halos is also defined at each redshift, according to the mass of the most massive substructure $M_2$ relative to that of the main halo $M_1$. In addition, we extract halos with $\MFOF>6.25\times10^{13}h^{-1}M_\odot$ at each redshift. Hence the number of the halos depends on redshift, and is indicated in Figure \ref{twod} and Table 2. Note that the virial overdensity $\Delta_{\rm vir} (z)$ depends on redshift; e.g., $\Delta_{\rm vir} (z=1)=203.2$.

Figure \ref{twod} (b), (c) and (d) show the results for halos at higher redshifts; $z=0.2$, 0.4 and 1, respectively. For each redshift, the PDF of OLS03 (blue curves) in the three panels are the same, but it slightly differs with redshift. The result for every redshift is basically similar to $z=0$. Also, all the histograms are well approximated by the beta distribution (Equation (\ref{betad})) and best-fit parameters are listed in Table 2.

According to Table 2, The mean value also has weak redshift dependence; it becomes smaller toward earlier redshifts. This is partly due to the fixed minimum mass for the sets of halos at different redshifts. At earlier redshifts, more massive fraction is chosen out of all the halos, so the mean axis ratio becomes smaller.

We have found that the values of mean and standard deviation of $a_1/a_2$ calculated directly from the simulation results agree within 5 \% from those in Table 2. In addition, the dependence of the mean value on mass scale and redshift are the same as discussed above, implying the goodness of the fitting by the beta distribution.

In addition, the statistical mass dependence of $a_1/a_2$ is also weak. Figure \ref{a1a2evol} shows the axis ratio $a_1/a_2$ of each halo against its $\Mvir$ (left panel) and $M_{500}$ (right panel). Except for the most massive part where the number of halos is small, the mass dependence of $a_1/a_2$ is even weaker than the three-dimensional axis ratio $A_1/A_3$ plotted in Figure \ref{eevol}. Therefore the minimum mass $\Mvir=6.25\times10^{13}h^{-1}M_\odot$ set in the above analysis is not so critical. The weak dependence of the PDF of $a_1/a_2$ on redshift, mass scales ($\Mvir$, $M_{500}$, $M_{2500}$) and the minimum mass of the halos are useful when the model is compared with observational data.

\begin{table}
\begin{center}
\begin{tabular}{c|c|cccc|cccc}
 & & \multicolumn{4}{c|}{all halos} & \multicolumn{4}{c}{single-halos only}\\
\hline
 & & $a$ & $b$ & mean & s.d. & $a$ & $b$ & mean & s.d. \\
\hline
$z=0$ & $\Mvir$ & 4.18 & 2.71 & 0.61 & 0.17 & 5.00 & 2.99 & 0.63 & 0.16 \\ 
$N=2004\times3$ & $M_{500}$ & 4.01 & 2.90 & 0.58 & 0.18 & 4.32 & 2.98 & 0.59 & 0.17 \\ 
 & $M_{2500}$ &4.35 & 3.39 & 0.56 & 0.17 & 3.92 & 3.14 & 0.56 & 0.17 \\  
\hline
$z=0.2$ & $\Mvir$ & 4.01 & 2.81 & 0.59 & 0.18 & 4.83 & 3.13 & 0.61 & 0.16 \\
$N=1550\times3$ & $M_{500}$ & 3.69 & 2.92 & 0.56 & 0.18 & 4.18 & 3.13 & 0.57 & 0.17 \\
 & $M_{2500}$ & 4.34 & 3.65 & 0.54 & 0.17 & 4.46 & 3.74 & 0.54 & 0.16 \\ 
\hline
$z=0.4$ & $\Mvir$ & 4.02 & 3.03 & 0.57 & 0.17 & 4.78 & 3.34 & 0.59 & 0.16 \\
$N=1101\times3$ & $M_{500}$ & 3.72 & 3.11 & 0.54 & 0.18 & 4.26 & 3.38 & 0.56 & 0.17 \\
 & $M_{2500}$ & 4.21 & 3.69 & 0.53 & 0.17 & 3.82 & 3.32 & 0.54 & 0.17 \\
\hline
$z=1$ & $\Mvir$ & 3.40 & 2.74 & 0.55 & 0.19 & 4.45 & 3.33 & 0.57 & 0.17 \\
$N=317\times3$ & $M_{500}$ & 3.22 & 2.83 & 0.53 & 0.19 & 3.89 & 3.25 & 0.54 & 0.17 \\
 & $M_{2500}$ & 3.93 & 3.80 & 0.51 & 0.17 & 3.95 & 3.77 & 0.51 & 0.17 \\ 
\end{tabular}
\end{center}
\caption{List of the parameters of the beta distribution (\ref{betad}) that approximates the PDF of projected axis ratio for $\Mvir$, $M_{500}$ and $M_{2500}$ for four redshifts. The mean $\mu=a/(a+b)$ and the standard deviation $\sigma=\sqrt{ab/[(a+b)^2(a+b+1)]}$ of the beta distribution are also shown.}
\end{table}

\subsection{Comparison with observational sample}

As an example of possible applications of our fitting formula for the PDF of projected axis ratio $a_1/a_2$, we attempt to compare it with the PDF for the observed halos estimated by \cite{Oguri10}. They measured the projected axis ratio from the weak lensing shear map of 18 clusters. In doing so, they assumed that the three-dimensional density distribution inside each halo follows a self-similar triaxial ellipsoid. The observed region of their clusters roughly corresponds to $M_{500}$, and their mean redshift is 0.23, so we compare their observation data with our model for $M_{500}$ of all the halos and $z=0.2$ in Table 2. The mean virial mass $\Mvir$ of our sample is roughly $2\times10^{14}h^{-1}M_\odot$.

The left panel of Figure \ref{oguri} plots the PDF of projected axis ratio of the observed halos in red symbols with error bars. \cite{Oguri10} compared this observational results with the PDF of OLS03 calculated by assuming $\Mvir=7\times 10^{14}h^{-1}M_\odot$ in Equation (\ref{JS}), corresponding to the mean mass of the observed clusters. We also plot the same PDF in the blue curve. In addition, our model is plotted in the black curve. The mean mass of our sample is smaller than that of \cite{Oguri10}, but this is not serious since the mass dependence of axis ratio $a_1/a_2$ is very weak (Figure \ref{a1a2evol}).

In order to include the possible effect of the observational uncertainty in $a_1/a_2$, \cite{Oguri10} convolved the PDF of OLS03 with the Gaussian function with $\sigma=0.15$, corresponding to the typical uncertainty of the measurement of the axis ratio (cf. Table 1 of \cite{Oguri10}). We repeated the same procedure for the PDF of OLS03 and our fitting formula.

The resulting PDFs are plotted in the right panel of Figure \ref{oguri}. Because of the large observational error bars, it is difficult to distinguish the PDF of OLS03 and our fitting formula. Our fitting formula is, however, based on the direct measurement of $a_1/a_2$ of the simulated halos, and therefore more reliable than that of OLS03 based on the self-similarity of halos. In the near future, precise observational data of numerous clusters will be provided by Subaru Hyper Suprime-Cam, for example. In order to perform much more precise comparison of observational data and theory, one should adopt the same estimator of the projected axis lengths both for observations and simulations. Hence more elaborate analyses will be a useful tool to test the non-sphericity of halos predicted in the CDM cosmology. Our current results are indeed towards such a purpose.

\section{Summary and Discussion}

We have studied the redshift evolution, and the mass- and radial-dependence of non-sphericity by analyzing halos from a cosmological N-body simulation. We approximated the density distribution of the halos by triaxial ellipsoids, employing the mass tensor $I=\sum xx$ (Equation (\ref{Ixx})). Our major findings are summarized as follows.
\begin{enumerate}
\item
For the first time, we compared the evolution of the axis lengths $A_k$ ($k=1$, 2, 3) of the triaxial ellipsoids of individual simulated halos with the prediction of the ellipsoidal collapse model (EC) {\it on the object-wise basis}. In general, the EC prediction roughly reproduces the simulation up to the turn-around epoch ($z\sim1$). After the turn-around epoch, however, the simulation substantially deviates from the EC prediction.

\item
The discrepancy in the statistical mass dependence of axis ratio $A_1/A_3$ at $z=0$ between the EC prediction and simulations has been reported in literature; massive halos are more spherical in EC, but those in simulations tend to be less spherical. For the first time, we demonstrated how and when the statistical mass dependence of $A_1/A_3$ of the simulated halos deviates from EC. While massive halos are more spherical initially, they gradually become less spherical after the turn-around epoch. This tendency is opposite to the EC prediction.

\item 
The averaged axis ratio $\langle A_1/A_3 \rangle$ over all the simulated halos has significant radial dependence as a function of enclosed mass $\Mel$. At $\Mel\gtrsim\MFOF$, $\langle A_1/A_3\rangle$ rapidly decreases due to filamentary structures around the halos. Inside $\Mel\sim\MFOF$, the radial dependence of $\langle A_1/A_3\rangle$ gradually changes with time; while $\langle A_1/A_3\rangle$ increases toward the inner region at $z=1$, it decreases at $z=0$. The radial dependence of $\langle A_1/A_3\rangle$ indicates that the halos are not necessarily self-similar (concentric, common axis ratio and orientation).

\item 
We examined how the probability distribution function (PDF) of $A_1/A_3$ of triaxial halos at $z=0$ depends on $\Mel$. The values of $\langle A_1/A_3\rangle$ at $\Mel=\MFOF/10$ is smaller than that of JS02 who employ the isodensity surface at $\rho=2500 \rho_{\rm crit}$ (roughly corresponding to the region of $\MFOF/10$) and adopt the self-similarity assumption. The difference is partly due to the different methods of triaxial modelling of halos; mass tensor and isodensity surface.

\item 
The projected axis ratio $a_1/a_2$ is a more relevant quantity to compare with observational data, and we find that it is sensitive to the self-similarity assumption as well. Therefore we calculated the PDF of $a_1/a_2$, not through those of three-dimensional $A_1/A_3$, but directly from the projected density distributions of the simulated halos for the first time. We provide a fitting formula for the resulting PDF, and show that the formula improves those in previous studies that assume the self-similarity of halos. Hence our fitting formula will play important roles in interpreting the future data of Subaru Hyper Suprime-Cam among others.
\end{enumerate}

Our current analysis is based on the dark-matter only simulation, which is reasonably justified in applying to weak-lensing observations. In a complementary fashion, X-ray observations of intracluster gas provide independent information on the non-sphericity of clusters. The density distribution of gas is, however, not necessarily identical to that of dark matter \citep{Lee03,Kawahara10}. Therefore, in order to study the non-sphericity of the gas density distribution without additional assumptions like hydrostatic equilibrium, hydrodynamical simulations including gas are needed. This is what we are currently working on, and will be presented elsewhere.

\clearpage

\section*{Acknowledgements}

We thank  an anonymous referee for many valuable comments. We gratefully acknowledge fruitful discussions with Benedikt Diemer, Hajime Kawahara, Surhud More, Masamune Oguri and S\'ebastien Peirani. This work is supported partly by JSPS Core-to-Core Program "International Network of Planetary Sciences", and by JSPS Grant-in-Aid for Scientific Research No. 26-11473 (D.S.), No. 25400236 (T. K.) and No. 24340035 (Y. S.). The numerical simulation in this work was carried out on Cray XC30 at Center for Computational Astrophysics, CfCA, of National Astronomical Observatory of Japan.

\clearpage

\begin{figure}[h]
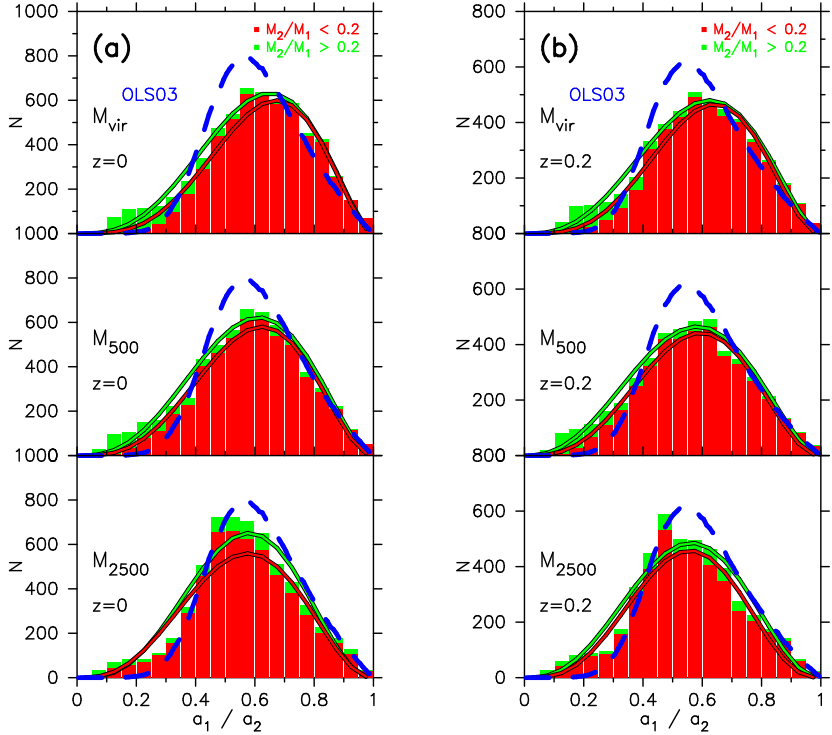

\begin{center}
\subfigure{
\FigureFile(50mm,50mm){pdf2d15.eps}
\qquad
\FigureFile(50mm,50mm){pdf2d13.eps}
}
\end{center}
\caption{PDF of the projected axis ratio at the three different mass
scale $\Mvir$ $M_{500}$ and $M_{2500}$, for four redshifts; $z=0$ (a),
0.2 (b), 0.4 (c), 1 (d). The solid curves show the best-fit beta
distributions (our model). For comparison, the PDF by OLS03 (based on
JS02) is also shown. The histogram is colored according to the
multiplicity of halos; $M_2/M_1<0.2$ (red), $M_2/M_1>0.2$ (green). The
multiplicity of halos ($M_2/M_1$) is determined separately at each
redshift. When calculating the PDF of OLS03,
$M=2\times10^{14}h^{-1}M_\odot$, corresponding to the mean mass of our
halos, is substituted in Equation (\ref{JS}). Only the simulated halos
with $\Mvir(z)>6.25\times10^{13}h^{-1}M_\odot$ are selected at each
redshift, and the number of halos is $3\times2004$ ($z=0$),
$3\times1550$ ($z=0.4$), $3\times1101$ ($z=0.2$), $3\times317$ ($z=1$).}
\label{twod}
\end{figure}

\addtocounter{figure}{-1}
\begin{figure}[h]
\begin{center}
\subfigure{
\FigureFile(50mm,50mm){pdf2d12.eps}
\qquad
\FigureFile(50mm,50mm){pdf2d09.eps}
}
\end{center}
\caption{Continued.}
\end{figure}

\begin{figure}[h]
\begin{center}
\subfigure{
\FigureFile(50mm,50mm){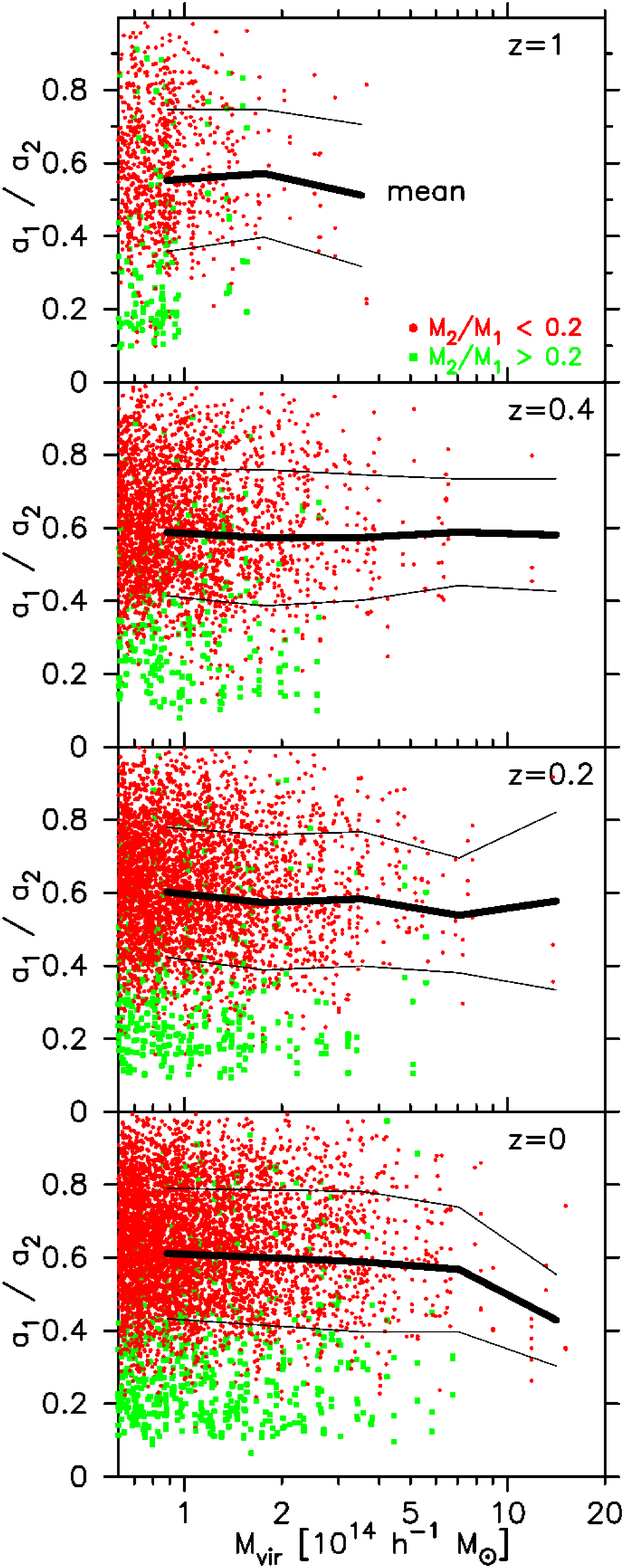}
\qquad
\FigureFile(50mm,50mm){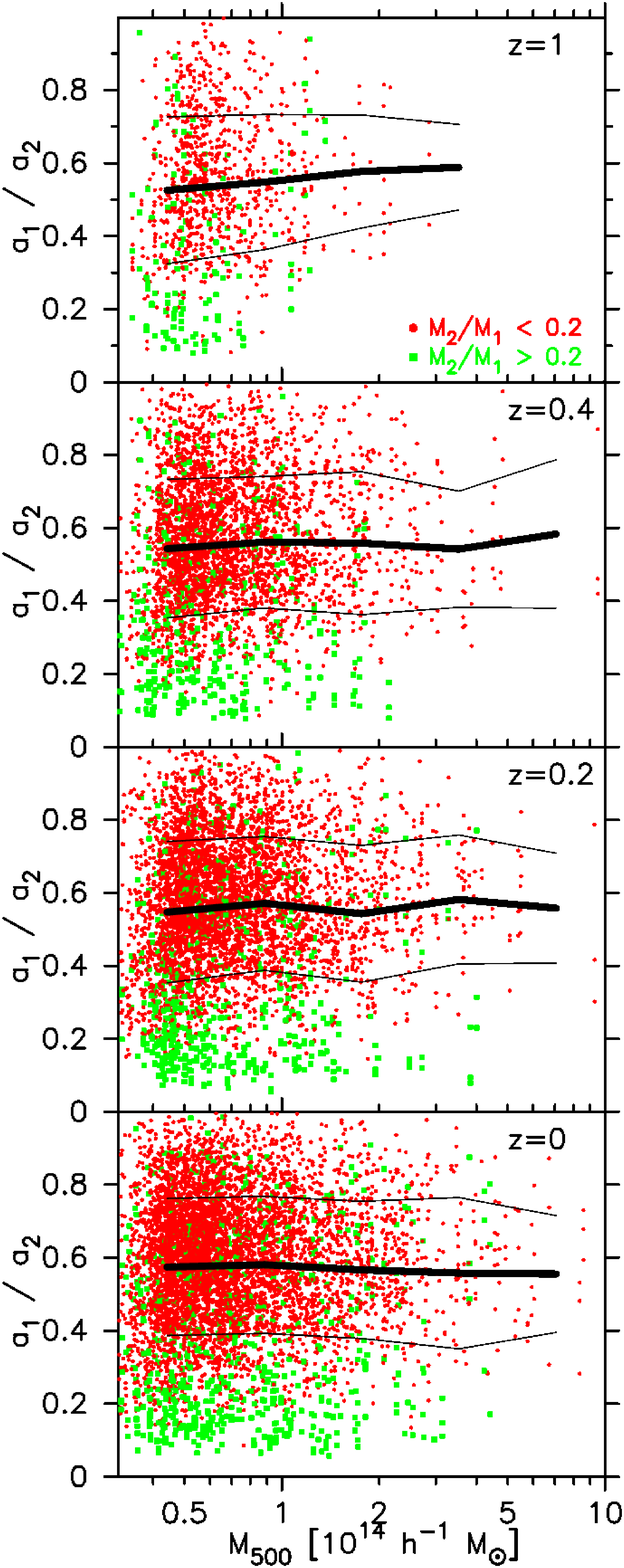}
}
\end{center}
\caption{Projected axis ratio $a_1/a_2$ of each halo at the four
different redshifts ($z=1$, 0.4, 0.2, 0) against its $\Mvir$ (left) and
$M_{500}$ (right). Each symbol indicates the result for each of the 2004
simulated halos; red circle are single-halos ($M_2/M_1<0.2$), green
squares are for multiple-halos ($M_2/M_1>0.2$). The thick and thin solid
lines indicate the mean and the standard deviation, respectively. The
halos are identified at each redshift, and their multiplicity
($M_2/M_1$) is also defined at each redshift. The number of the
simulated halos is $3\times2004$ ($z=0$), $3\times1550$ ($z=0.4$),
$3\times1101$ ($z=0.2$), $3\times317$ ($z=1$).}  \label{a1a2evol}
\end{figure}

\begin{figure}[h]
\begin{center}
\subfigure{
\FigureFile(50mm,50mm){oguri.eps}
\qquad
\FigureFile(50mm,50mm){oguriG.eps}
}
\end{center}
\caption{Comparison of PDFs of projected axis ratio $a_1/a_2$. The red
symbols with error bars show the results from the 18 clusters in the
weak lensing analysis by \cite{Oguri10}. The PDF of OLS03 is plotted in
the blue dashed curve. Following \cite{Oguri10}, we use
$\Mvir=7\times10^{14}h^{-1}M_\odot$, corresponding to the mean mass of
the observed clusters, when calculating the PDF of OLS03 through
Equation (\ref{JS}). The black solid curve indicate our fitting formula
for the PDF of $a_1/a_2$ at $M_{500}$ of all the halos at $z=0.2$ (Table
2). For the PDF of OLS03 and ours, the left panel illustrates the
original PDFs, while the right panel shows those convolved with the
Gaussian function with $\sigma=0.15$, corresponding to the typical
uncertainty for $a_1/a_2$ in the lensing analysis (cf. Figure 3 and
Table 1 of \cite{Oguri10}).}  \label{oguri}
\end{figure}

\end{document}